\documentclass[11pt]{article} 

\usepackage[utf8]{inputenc} 

\usepackage{geometry} 
\usepackage{graphicx} 

\usepackage{booktabs} 
\usepackage{array} 
\usepackage{verbatim} 
\usepackage{caption}
\usepackage{subcaption}
\usepackage{hyperref}

\usepackage{fancyhdr} 
\usepackage{sectsty}
\allsectionsfont{\sffamily\mdseries\upshape} 

\usepackage[nottoc,notlof,notlot]{tocbibind} 
\usepackage[titles,subfigure]{tocloft} 

\usepackage{amsmath}
\usepackage{amssymb}

\def\SHAPE#1{\mbox{{\tt #1}}}

\title{Spaced seeds improve $k$-mer-based metagenomic classification}
\author{Karel  B\v{r}inda \and Maciej Sykulski \and Gregory
  Kucherov\thanks{corresponding author, Gregory.Kucherov@u-pem.fr}}
\date{\small Laboratoire d'Informatique Gaspard-Monge, Universit\'e Paris-Est\&CNRS, France} 

\begin{document}
\maketitle

\begin{abstract}
Metagenomics is a powerful approach to study genetic content of
environmental samples, which has been strongly promoted by NGS
technologies. To cope with massive data involved in modern metagenomic
projects, recent tools \cite{pmid23828782,pmid24580807} rely on the
analysis of $k$-mers shared between the read to be classified and
sampled reference genomes. Within this general framework, we show in
this work that {\em spaced seeds} provide a significant improvement of
classification accuracy as opposed to traditional {\em
  contiguous} $k$-mers. We support this thesis through a series a
different computational experiments, including simulations of
large-scale metagenomic projects. 

Scripts and programs used in this study, as well as 
supplementary material, are available from 
\href{http://github.com/gregorykucherov/spaced-seeds-for-metagenomics}{http://github.com/gregorykucherov/spaced-seeds-for-metagenomics}.

\end{abstract}

\section{Introduction}

Metagenomics is a powerful approach to study genetic
material contained in environmental samples. The advent of high-throughput sequencing
technologies ({\em Next-Generation Sequencing}, commonly abbreviated
to NGS) revolutionized this approach, by avoiding the need of cloning the DNA and thus 
greatly facilitating the obtention of metagenomic samples, at the same
time drastically decreasing its
price. Early examples of metagenomic projects include the analysis of
samples of seawater \cite{pmid15001713,pmid22028628}, human
gut \cite{QinEtAlNature10}, or soil
\cite{VogelEtAlNatRevMicro09}. 
Present-day metagenomic studies focus on various bacterial, fungal or viral
populations, exemplified by the
Human Microbiome project \cite{pmid19819907} that investigates microbial communities at
different sites of human body. 

Modern metagenomics deals with vast sequence datasets. On the one
hand, metagenomic samples ({\em metagenomes}) obtained through NGS
are commonly measured by tens or even hundreds
of billions of bp \cite{NatMeth09}. These sequences generally come
from a number of different species, some of which either have a previously sequenced
reference genome, or have a related sequenced species sufficiently close phylogenetically to determine this
relationship by sequence comparison. Other sequences, however, may
come from organisms that have no sufficiently close relatives with
sequenced genomes, or from DNA fragments that show no significant
similarity with any available genomic sequence. The {\em metagenomic classification} problem is to
assign each sequence of the metagenome to a corresponding taxonomic
unit, or to classify it as 'novel'. 

A way to improve the accuracy of metagenomic classification is to
match the me\-ta\-ge\-nome against as large set of known genomic sequences
as possible. With many thousands of completed microbial genomes
available today, modern metagenomic projects match their samples
against genomic databases of tens of billions of bp
\cite{pmid24580807}. 

{\em Alignment-based classifiers} \cite{pmid22962338} proceed by 
aligning metagenome sequences to each of the known genomes from the
reference database, in order to use the best alignment score as
an estimator of the phylogenetic ``closeness'' between the sequence
and the genome. This could be done with generic alignment program,
such as {\sc Blast} \cite{GBLAST97}, {\sc Blastz} \cite{BLASTZ03}, or
{\sc Blat} \cite{BLAT02}. While this approach can be envisaged for small datasets
(both metagenome and database) and is actually used in such software tools as {\sc
  Megan} \cite{pmid21690186} or {\sc PhymmBL} \cite{pmid21527926} (see
\cite{pmid22962338} for more), it is unfeasible on the scale of modern
metagenomic projects. 
On the other hand, there exists a multitude of
specialized tools for
aligning NGS reads -- {\sc BWA} \cite{pmid19451168}, {\sc Novoalign}
(\url{http://www.novocraft.com/}), 
{\sc GEM} \cite{pmid23103880}, {\sc Bowtie} \cite{pmid19261174},
just to mention a few popular ones -- which perform alignment at a
higher speed and are adjusted to specificities of NGS-produced
sequences. Still, aligning multimillion read sets against thousands of
genomes remains computationally difficult even with optimized
tools. Furthermore, read alignment algorithms are usually designed to
compute high-scoring alignments only, and are often unable to report
low-quality alignments. 
As a result, a large fraction of  reads may remain unmapped \cite{LindgreenEtAl15}.

Several techniques exist to reduce the computational complexity of
this approach. One direction is to pre-process the metagenomic sample
in order to assemble reads into longer contigs, potentially improving
the accuracy of assignment. Assembly of metagenomic reads has been a subject of many works 
(see \cite{pmid22966151}) and remains a fragile approach, due to
its error-proneness and high computational complexity. Overall, it
appears feasible mostly for small-size projects with relatively high
read coverage. 

To cope with increasingly large metagenomic projects,
alignment-free methods have recently come into use.
Those methods do not compute read alignments, thus do not come with
benefits of these, such as gene identification. 
Alignment-free
sequence comparison is in itself an established research area,
reviewed in a recent dedicated special issue \cite{pmid24819825}. Most
of alignment-free comparison methods are based on the analysis of
words, usually of fixed size  ($k$-mers), occurring in input
sequences. A popular approach is to compute the distance between
{\em frequency vectors} of all $k$-mers in each of the sequences. In
the context of metagenomics, however, when one of the sequences is
short (NGS read), the analysis is based on the shared $k$-mers, without
taking into account their multiplicities in reference genomes. This is also dictated by the
prohibitive computational cost of computing and storing $k$-mer multiplicities
for metagenomics-size data. 

Two recently released tools -- {\sc LMAT} \cite{pmid23828782} and {\sc
  Kraken} \cite{pmid24580807} -- perform metagenomic classification
of NGS reads based on the analysis of shared $k$-mers between an input read and each
genome from a pre-compiled database. Given a taxonomic tree involving
the species of the database, those tools ``map'' each read to a node
of the tree, thus reporting the most specific taxon or clade the read is
associated with. Mapping is done by sliding through all $k$-mers
occurring in the read and determining, for each of them, the genomes
of the database containing the $k$-mer. Based on obtained counts
and tree topology, algorithms \cite{pmid23828782,pmid24580807} assign
the read to the tree node ``best explaining'' the counts. 
Further similar tools have been published during last months \cite{pmid25879410,pmid25884504}.

The goal of this work is to show that the metagenomics classification
based on the analysis of shared $k$-mers can be improved by using {\em
  spaced $k$-mers} rather than contiguous $k$-mers. 

The idea of using spaced $k$-mers goes back to the concept of {\em
  spaced seeds} for {\em seed-and-extend} sequence comparison
\cite{PatternHunter02,BurkhardtKarkkainen03}. There, the idea 
is to use as a seed (i.e. local match triggering an
alignment) a sequence of matches interleaved with ``joker positions''
holding either matches or mismatches. The pattern specifying the
sequence of matches and jokers is called the spaced seed. Remarkably,
using spaced seeds instead of contiguous seeds significantly improves
the sensibility-selectivity trade-off with almost no incurred
computational overhead. This has been first observed
in \cite{PatternHunter02} and then extended and formally analysed in
a series of further works, see
\cite{KucherovNoeRoytbergJBCB06,BrownBA08} and references therein. 

Recently, it has been reported in several works that spaced seeds
bring an improvement in alignment-free comparison as well. In
\cite{DBLP:journals/bioinformatics/LeimeisterBHLM14}, it is shown that
comparing frequency vectors of {\em spaced $k$-mers} ($k$-mers
obtained by sampling must-match positions of one or several spaced
seeds), as opposed to contiguous $k$-mers, leads to a more accurate estimation of phylogenetic distances
and, as a consequence, to a more accurate reconstruction of phylogenetic
trees. 
In \cite{pmid25685176}, the authors studied another measure -- the
number of pairs of matching (not necessarily aligned) spaced $k$-mers between the input sequences
-- and showed that it provides an even better estimator of the
phylogenetic distance. 
In \cite{pmid25393923}, it is shown that the number of hits of
appropriately chosen spaced seeds  in {\em aligned} sequences 
and their {\em coverage}
(i.e. the total number of matched positions covered by all hits) provides a much
better estimator for the alignment distance than the same measures
made with contiguous seeds. From a machine learning perspective,
works \cite{OnoderaShibuya13,pmid25033408} show that spaced seeds provide better
string kernels for SVM-based sequence classification, confirmed by
experiments with protein classification (see also
\cite{pmid25393923}). 

In this work, we show that using spaced $k$-mers significantly
improves the accuracy  of metagenomic classification of NGS reads as
well.  
To support this thesis, we consider different scenarios. As a test
case, we first
study the problem of discriminating a read between two genomes,
i.e. determining which of the two genomes is ``phylogenetically closer''
to the read. We then analyse the correlation between the quality of an
alignment of a read to a genome and the seed count for this read,
defined either as the number of hits or as the coverage. This analysis 
provides an insight into how well one can estimate the similarity between
a read and a genome out of $k$-mer occurrences. Finally, we make a
series of large-scale metagenomic classification experiments with
{\sc Kraken} software \cite{pmid24580807} extended by the possibility
of dealing with spaced seeds. 
These experiments demonstrate an improved classification
accuracy at the genus and family levels caused by the use of spaced seeds instead of contiguous ones.

\section{Preliminaries}

A spaced seed is a binary pattern over symbols \SHAPE{\#} and
\SHAPE{-} denoting match and joker respectively. For example, seed
\SHAPE{\#-\#\#} specifies a match followed by either match or mismatch
followed by two consecutive matches. A seed acts as a mask for
comparing short oligonucleotides, for example sequences {\tt gaat} and {\tt gcat}
differ at the 2nd position, but they match via seed \SHAPE{\#-\#\#} as
the 2nd position is masked out. The number of \SHAPE{\#}'s in a
seed, called {\em weight}, defines the number $k$ of matching
nucleotides. In the above example, $k=3$ and the matching (spaced)
$k$-mer is {\tt gat}. 
In a slightly different terms, a seed is a pattern that specifies a
small part of a gapless alignment seen as a binary sequence of matches
and mismatches. For example, seed \SHAPE{\#-\#\#} occurs in (or {\em
  hits}, as usually said) any alignment containing a match
followed by another two matches shifted by one position. 

When spaced seeds are used for sequence alignment within the {\em
  seed-and-extend paradigm} \cite{BrownBA08}, a pair of matching
$k$-mers (or, sometimes, a matching sequence of several closely located 
$k$-mers) indicates a potential alignment of interest in which the two
$k$-mers are aligned together. When spaced seeds are used for
alignment-free sequence comparison \cite{pmid24819825}, the goal is to estimate the
similarity of two sequences based on the {\em number} of matching
$k$-mers, with no or limited information about their positions in one
or both sequences. This measure can be formalized in several different ways.

In the
context of metagenomic classification of NGS reads, the goal is
to estimate the evolutionary distance between a short read and a long sequence
(genome),  which can be modeled by the best alignment score of
the read against the sequence. Due to large genome size and
large number of reference genomes involved in computations, one of the
constraints is to avoid 
keeping track of positions of $k$-mers in genomes. We can only afford
constructing an index to quickly answer queries whether or not a $k$-mer
occurs in a given genome, without information on its
positions. On the other hand, $k$-mer positions in the query read can
be included to the analysis. Therefore, we have to derive our
estimation from the number of $k$-mers shared between the read and the
genome, together with their positions in the read alone. 

One simple estimator is the number of $k$-mers in the read that occur
in the target genome\footnote{Here identical $k$-mers occurring at distinct
  positions in the read are considered distinct.}, we call this
measure the {\em hit number}. However, one may
want to favor cases when matching $k$-mers cover a larger part of the
read, vs. those with matching $k$-mers clumped together due to
overlaps. This leads to the concept of {\em coverage} \cite{pmid25393923},
earlier used in seed-based alignment algorithms as well
\cite{NoeKucherovBMC04,DBLP:conf/spire/BensonM08}. 
The coverage is the total number of positions covered by all matching
$k$-mers. 
For example,
consider seed \SHAPE{\#-\#\#} and read \SHAPE{gaatcagat}. Assume the
seed hits at positions 1,4,6, i.e. $k$-mers \SHAPE{g-at},
\SHAPE{t-ag} and \SHAPE{a-at} occur in the
target genome (joker symbol is shown for the sake of clarity). Here,
the hit number is 3, and the coverage is 7 as seven positions are
covered by hits, namely positions 1,3,4,6,7,8,9. Hit number and
coverage are two estimators studied in this work. 

\section{Results}

\subsection{Binary classification}
\label{binary}

From the machine learning viewpoint, hit number and coverage 
can be viewed as instances of kernel functions for sequence data (see
e.g. \cite{pmid18974822}). Our first step was to compare their
capacity w.r.t. a simple binary classification task. Assume the
(impractical) case when our database contains only two genomes. Given
a read, we have to decide which of the two genomes the read is closer
to. How good can we be at that? Which kernel works better for this
task? And are spaced seeds better than contiguous seeds here? 

\subsubsection{Classifying aligned reads}
\label{seed-count}

As mentioned in Introduction, the ``distance'' of a read
to a genome translates naturally to the score of the best alignment of
the read to the genome.  Given two genomes, we want to tell which of
them is closer to a given read. 

Consider alignments $A_s,A_l$ of a random read to the two genomes, and assume
they are gapless and have mismatch probabilities $p_s,p_l$ respectively, with
$p_s<p_l$. Throughout this paper the read length is set to $100$, a
typical length of {\sc Illumina} read. Therefore, the alignments can be
thought of as random binary strings of length $100$ of matches and mismatches, with
mismatch probabilities $p_s$ and $p_l$ respectively. Given a seed, we compute two counts
$C_s$ and $C_l$ on alignments
$A_s$ and $A_l$ respectively, where by 'count' we mean, unless otherwise stated,
either the hit number or the coverage. For example, if the seed is
\SHAPE{\#-\#\#} and the alignment \SHAPE{111101111} (\SHAPE{1}
stands for match and \SHAPE{0} for mismatch), then the hit number is
3 and the coverage is 7. Note that in this model, common (spaced)
$k$-mers are assumed to occur necessarily at the same position in the
read alignment, although in reality, a $k$-mer of the read may not be
aligned with the same $k$-mer in the genome. However, in this first
experiment, we abstract from this fact. 

If $C_s>C_l$ (resp. $C_s<C_l$), then we report a correct
(resp. incorrect) classification, otherwise a tie is reported. 
By iterating this computation, we estimate the
probability of correct/incorrect classification
for each parameter set. 

\begin{table}
\centering
\begin{subtable}{\textwidth}
\centering
\caption{$p_s=0.05$, $p_l=0.1$\label{00501}}
\begin{tabular}{|c|c|c|c|c|}
\hline
&\multicolumn{2}{|c|}{contiguous}&
\multicolumn{2}{|c|}{spaced}\\
\hline
weight&Hits&Cover&Hits&Cover\\
\hline
16 & .85/.14 & .85/.14 & .87/.11 & .88/.11 \\
18 & .84/.14 & .85/.14 & .87/.12 & .88/.11 \\
20 & .83/.15 & .84/.15 & .87/.12 & .87/.12 \\
22 & .82/.16 & .83/.15 & .86/.12 & .87/.12 \\
24 & .80/.16 & .81/.15 & .85/.12 & .87/.12 \\
\hline
\end{tabular}
\end{subtable}
\medskip

\begin{subtable}{\textwidth}
\centering
\caption{$p_s=0.1$, $p_l=0.2$\label{0102}}
\begin{tabular}{|c|c|c|c|c|}
\hline
&\multicolumn{2}{|c|}{contiguous}&
\multicolumn{2}{|c|}{spaced}\\
\hline
weight&Hits&Cover&Hits&Cover\\
\hline
16 & .86/.09 & .87/.09 & .93/.05 & .94/.05 \\
18 & .81/.08 & .81/.08 & .91/.05 & .92/.05 \\
20 & .74/.07 & .74/.07 & .89/.05 & .90/.05 \\
22 & .65/.06 & .65/.06 & .85/.05 & .86/.05 \\
24 & .55/.04 & .56/.04 & .79/.04 & .81/.05 \\
\hline
\end{tabular}
\end{subtable}
\medskip

\begin{subtable}{\textwidth}
\centering
\caption{$p_s=0.2$, $p_l=0.3$\label{0203}}
\begin{tabular}{|c|c|c|c|c|}
\hline
&\multicolumn{2}{|c|}{contiguous}&
\multicolumn{2}{|c|}{spaced}\\
\hline
weight&Hits&Cover&Hits&Cover\\
\hline
16 & .40/.06 & .40/.06 & .63/.07 & .64/.07 \\
18 & .28/.04 & .28/.03 & .50/.05 & .50/.05 \\
20 & .18/.02 & .18/.02 & .37/.03 & .37/.03 \\
22 & .12/.01 & .12/.01 & .25/.02 & .26/.02 \\
24 & .08/.00 & .08/.00 & .17/.01 & .18/.01 \\
\hline
\end{tabular}
\end{subtable}
\caption{\small Each entry contains a pair ``Probability of correct
  classification/Probability of incorrect classification''. The
  remaining fraction estimates the probability of a tie. Spaced seeds used for hit number: 
  \SHAPE{\#\#\#-\#\#-\#-\#---\#-\#-\#--\#--\#\#\#\#\#} (weight 16),
  \SHAPE{\#\#\#\#-\#\#--\#\#--\#--\#-\#-\#-\#\#-\#\#\#\#} (weight 18),
\SHAPE{\#\#\#\#\#-\#\#\#---\#\#--\#--\#-\#-\#-\#\#-\#\#\#\#} (weight 20), 
\SHAPE{\#\#\#\#\#\#--\#\#\#\#----\#--\#\#-\#-\#--\#\#\#\#\#\#\#} (weight 22),
\SHAPE{\#\#\#\#\#\#\#--\#\#\#\#----\#--\#\#-\#-\#--\#\#\#\#\#\#\#\#} (weight 24). 
Spaced seeds used for coverage: 
  \SHAPE{\#\#\#-\#\#\#--\#-\#--\#-\#\#--\#\#\#\#\#} (weight 16),
  \SHAPE{\#\#\#-\#--\#\#\#-\#\#--\#-\#--\#\#\#-\#\#\#\#} (weight 18),
\SHAPE{\#\#\#-\#-\#\#-\#-\#\#--\#\#--\#\#\#-\#-\#\#\#\#\#} (weight 20), 
\SHAPE{\#\#\#\#-\#\#-\#-\#-\#\#-\#\#\#-\#-\#--\#\#--\#\#\#\#\#} (weight 22),
\SHAPE{\#\#\#\#-\#\#-\#-\#-\#\#-\#\#\#-\#-\#---\#\#\#--\#\#\#\#\#\#} (weight 24).  
}
\end{table}


The results are presented in Tables~\ref{00501},\ref{0102},\ref{0203}. 
 In all cases, spaced seeds show a
better classification power. In some cases, the
difference is striking: for example, if we want to discriminate between alignments
with mismatch probabilities $0.1$ and $0.2$ (Table~\ref{0102}) using
seeds of weight $22$, then a spaced seed yields $86\%$ of correct
classifications 
(coverage count), whereas the contiguous seed
correctly classifies only $65\%$ of cases, 
whereas the fraction of incorrect classifications is essentially
the same. 
The results also show a slight edge of the coverage count over the hit
number, which suggests the superiority of the latter that will be
confirmed later on in other experiments. 

\subsubsection{Classifying unaligned reads}
\label{binary-on-genomes}

Let us now turn to a more practical setting, where we want to classify
reads coming from a genome $G$ between two other genomes $G_1$ and
$G_2$ based on the phylogenetic closeness. 

To study this, we implemented the
following experimental setup. 
Using {\sc dwgsim} read simulator (\url{https://github.com/nh13/DWGSIM}), we generate single-end
{\sc Illumina}-like reads from genome $G$. In all experiments, we assumed
1\% of base mutations (substitutions only), and 2\% of sequencing
errors ({\sc dwgsim} options \texttt{-e 0.02 -r 0.01 -R 0}). Given a
seed, (contiguous or spaced) $k$-mers of $G_1$ and $G_2$ are indexed
to support existence queries only. For each read, all $k$-mers are
queried against $G_1$ and $G_2$ and corresponding counts $C_1$ and
$C_2$ are computed. If $C_1>C_2$ (resp. $C_1<C_2$), the read is classified to be closer
to $G_1$ (resp. $G_2$), otherwise a tie is reported. 
Besides considering
absolute counts (hit number and coverage), we also considered hit
number normalized by the number of distinct $k$-mers in the corresponding
genome (computed at the indexing stage). This measure approximates
the Jaccard index \cite{pmid21810899} and reflects the Bayesian probability of seeing a
$k$-mer relative to the ``$k$-mer-richness'' of genomes.

Note that the counts are here computed relative to the
whole genome, as it is done in the approach of
\cite{pmid23828782,pmid24580807} (see Introduction). This means that the $k$-mers
occurring in the read are looked up in the whole genome, without
guarantee, however, that these $k$-mers are closely located in the
genome and contribute to the same read alignment. This makes the seed
weight an important parameter, as seeds of low weight may result in a
high read count which does not evidence any alignment of the read, due
to random character of the $k$-mer hits. 

We experimented with bacterial genomes belonging to {\em Mycobacterium}
genus. Members of this genus present low interspecies genetic
variability and their phylogeny remains uncertain
\cite{pmid21684354}. 

If $G$ coincides with one of $G_1,G_2$, i.e. reads have to be
classified between its source genome and another genome, then all
estimators correctly classify nearly all reads as soon as $G_1$ and
$G_2$ are genomes of distinct species. For example, classifying reads obtained
from {\em Mycobacterium tuberculosis} (H37Rv, acc. NC\_018143) against {\em M. tuberculosis} itself
and {\em M.avium} (104, NC\_008595) led to more than 99\% of correct classifications for
all estimators. 


The case when $G$ is distinct from $G_1,G_2$ appears more
interesting. It corresponds to the real-life situation when reads to
be classified can come from a genome that is not represented in the database. 
Here we expect our procedure to determine whether $G$ is
phylogenetically closer to $G_1$ or to $G_2$. 

For example, we classified {\em M.vanbaalenii} (PYR-1, NC\_008726) reads against
{\em M.smeg\-matis} (MC2 155, NC\_018289) and {\em M.gilvum} (PYR-GCK, NC\_009338) genomes. 
Alternative phylogenies given in \cite[Fig.1-4]{pmid21684354} imply different
evolutionary relationships among these three species. Our results, shown in
Table~\ref{smeg_gil_van}, suggest that  {\em M.vanbaalenii} is closer
to  {\em M.gilvum} than to  {\em M.smegmatis}. For non-normalized hit
number and coverage estimators, this conclusion is supported by seeds of weight 16 or more, while
weight 14 supports the opposite conclusion. This is due to
spurious hits that become dominating when the weight drops to 14, and
to the larger size of {\em M.smegmatis} genome (6.99Mbp) compared to
{\em M.gilvum} (5.62Mbp). 
This effect is corrected by Jaccard index due to normalization by the
number of distinct $k$-mers (6.09M for {\em M.smegmatis}
vs. 4.96M for {\em M.gilvum} for the spaced seed of weight 14). Overall, we observe
a significantly sharper discrimination produced by spaced seeds
compared to contiguous seeds. 

\begin{table}
\caption{Classification of {\em Mycobacterium vanbaalenii} reads
  against {\em Mycobacterium
    smegmatis} and {\em Mycobacterium gilvum} genomes. Each entry contains a pair ``Fraction (in \%) of reads
  classified closer to {\em M.smegmatis}\, /\,Fraction of reads
  classified closer to {\em M.gilvum}''. 
\label{smeg_gil_van}}
\centering
\begin{tabular}{|r|c|c|c|c|c|}
\hline
& \multicolumn{5}{c|}{weight}\\
\cline{2-6}
                        & 14                  & 16                   & 18                   & 20                   & 22\\
\hline
  contig hit nb    & 52/41 & 39/48 & 24/37 & 11/24 & 07/17\\
  contig cover  & 54/42 & 44/47 & 25/37 & 11/24 & 06/17\\
 contig Jaccard & 30/70 & 35/61 & 23/42 & 11/26 & 06/18\\
\hline
  spaced hit nb    & 51/40 & 34/47 & 20/40 & 12/32 & 08/23\\
  spaced cover & 53/42 & 39/51 & 21/42 & 12/32 & 08/25\\
  spaced Jaccard & 28/72 & 32/66 & 20/50 & 11/33 & 08/27\\
\hline
\end{tabular}
\end{table}

\begin{table}
\caption{Classification of {\em Bacillus thuringiensis} reads
  against {\em Bacillus anthracis} and {\em Bacillus cereus} genomes. Each entry contains a pair ``Fraction (in \%) of reads
  classified closer to {\em B.anthracis}\, /\,Fraction of reads
  classified closer to {\em B.cereus} ''. \label{thur_anth_cer}}
\centering
\begin{tabular}{|r|c|c|c|c|c|}
\hline
& \multicolumn{5}{c|}{weight}\\
\cline{2-6}
                        & 14            & 16         & 18        & 20        & 22\\
\hline
  contig hit nb    & 83/14 & 81/11 & 79/09 & 77/08 & 76/08\\
  contig cover     & 78/17 & 80/12 & 79/09 & 77/08 & 76/08\\
  contig Jaccard & 87/13 & 87/11 & 85/09 & 83/08 & 82/08 \\
\hline
  spaced hit nb    & 83/13 & 82/11 & 80/09 & 79/09 & 79/08\\
  spaced cover    & 80/15 & 81/11 & 80/09 & 79/09 & 79/08\\
  spaced Jaccard & 88/12 & 88/11 & 85/09 & 84/08 & 84/08 \\
\hline
\end{tabular}
\end{table}

We also performed a series of experiments with the large and genetically variable  {\em Bacillus}
genus. Table~\ref{thur_anth_cer} shows a demonstrative experiment with
members of {\em Bacillus cereus} group: {\em B.thuringiensis} (serovar
konkukian 97-27, NC\_005957), {\em B.anthracis} (Ames, NC\_003997), and {\em B.cereus}
(ATCC 14579, NC\_004722). The three bacteria are close to the point of being
considered to be different lineages of a single {\em B.cereus} species
\cite{pmid10831447}. 
The results provide a strong support that
the {\em B.thuringiensis}  strain is closer to the {\em B.anthracis}
strain than to the {\em
  B.cereus} strain, which agrees with phylogenies reported in the literature
\cite{pmid20504335}. Indeed, the {\em B.thuringiensis} strain 
and the {\em B.anthracis} strain 
have a much higher pairwise identity rate
than the former has with the {\em B.cereus} strain 
(estimated
DNA-DNA hybridization distance 81\% vs. 45\%, as computed by {\sc
  GGDC} \cite{pmid21304684}). 

\begin{table}
\caption{Classification of {\em Bacillus licheniformis} reads
  against {\em Bacillus anthracis} and {\em Bacillus pumilus} genomes. Each entry contains a pair ``Fraction (in \%) of reads
  classified closer to {\em B.anthracis}\, /\,Fraction of reads
  classified closer to {\em B.pumilus} ''. 
\label{anth_pum_lich}}
\centering
\begin{tabular}{|r|c|c|c|c|c|}
\hline
& \multicolumn{5}{c|}{weight}\\
\cline{2-6}
                        & 14            & 16         & 18        & 20        & 22\\
\hline
  contig hit nb    & 47/40 & 24/25 & 04/07 & 0.9/3.8 & 0.3/2.4\\
  contig cover     & 49/46 & 24/25 & 04/07 & 0.9/3.8 & 0.3/2.4\\
  contig Jaccard  & 37/62 & 24/30 & 04/08 & 0.8/4.0 & 0.3/2.6\\
\hline
  spaced hit nb    & 49/33 & 22/27 & 04/10 & 1.0/5.4 & 0.4/3.3\\
  spaced cover    &  53/39 & 23/27 & 04/10 & 1.0/5.4 & 0.5/3.3\\
  spaced Jaccard &  41/59 & 22/35 & 04/10 & 0.9/4.6 & 0.4/3.5\\
\hline
\end{tabular}
\end{table}

However, for species with low sequence similarity, a large
majority of reads may have no hits to either genome, and only a small
fraction of reads may reveal a significant difference in
distance. This situation is illustrated in Table~\ref{anth_pum_lich}
showing the results of classification of {\em B.licheniformis}
(ATCC 14580, NC\_006270) reads against {\em B.anthracis} (Ames, NC\_003997)
and {\em B.pumilus} genomes (SAFR-032, NC\_009848). The results support a higher similarity of {\em
  B.licheniformis} to {\em B.pumilus} than to {\em B.anthracis}, but
the difference is revealed on a very small fraction of reads. 
The conclusion, however, is significant 
as those reads represent the
majority of reads having any hits to one of the genomes. 
As with the previous example, this result is
confirmed by previously reported phylogenies \cite{pmid20504335}.  


In all our experiments reported in this section, spaced seeds
showed a better classification capacity. 
The difference is especially significant in ``nontrivial'' cases involving relatively
dissimilar genomes, such as those illustrated by 
Tables~%
\ref{smeg_gil_van} and \ref{anth_pum_lich}. 
While the difference between hit number and coverage estimators
appeared insignificant (in agreement with results of
Section~\ref{seed-count}), the Jaccard index generally provides a
more distinct discrimination and, combined with a spaced seed, appears
to be the best estimator.

\subsection{Correlation of counts with alignment quality}
\label{seed-meta}

\begin{figure}
\centering
\begin{subfigure}[b]{0.49\textwidth}
  \includegraphics[width=\textwidth]{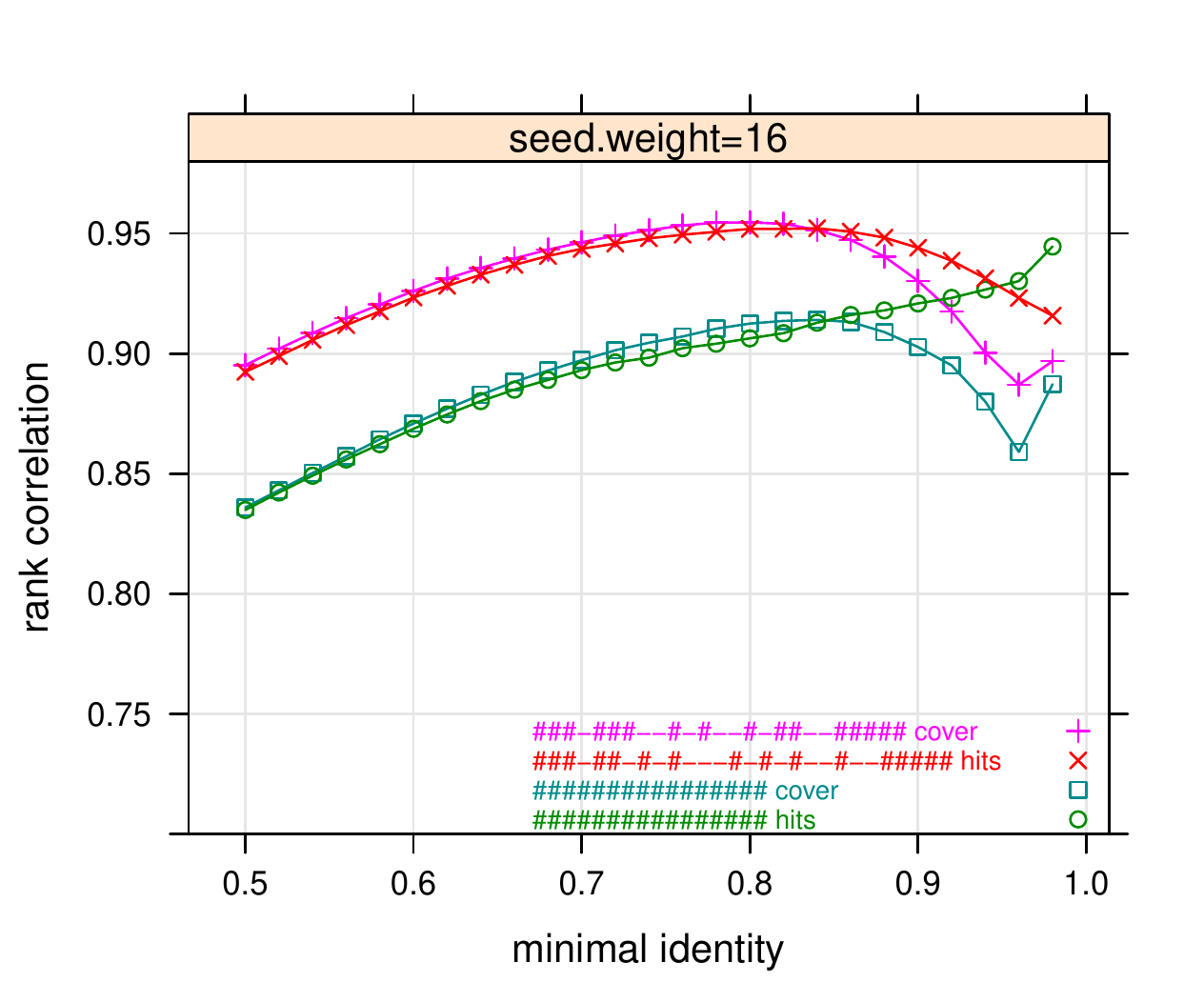}
                \label{weight16}
              \end{subfigure}
\begin{subfigure}[b]{0.49\textwidth}
  \includegraphics[width=\textwidth]{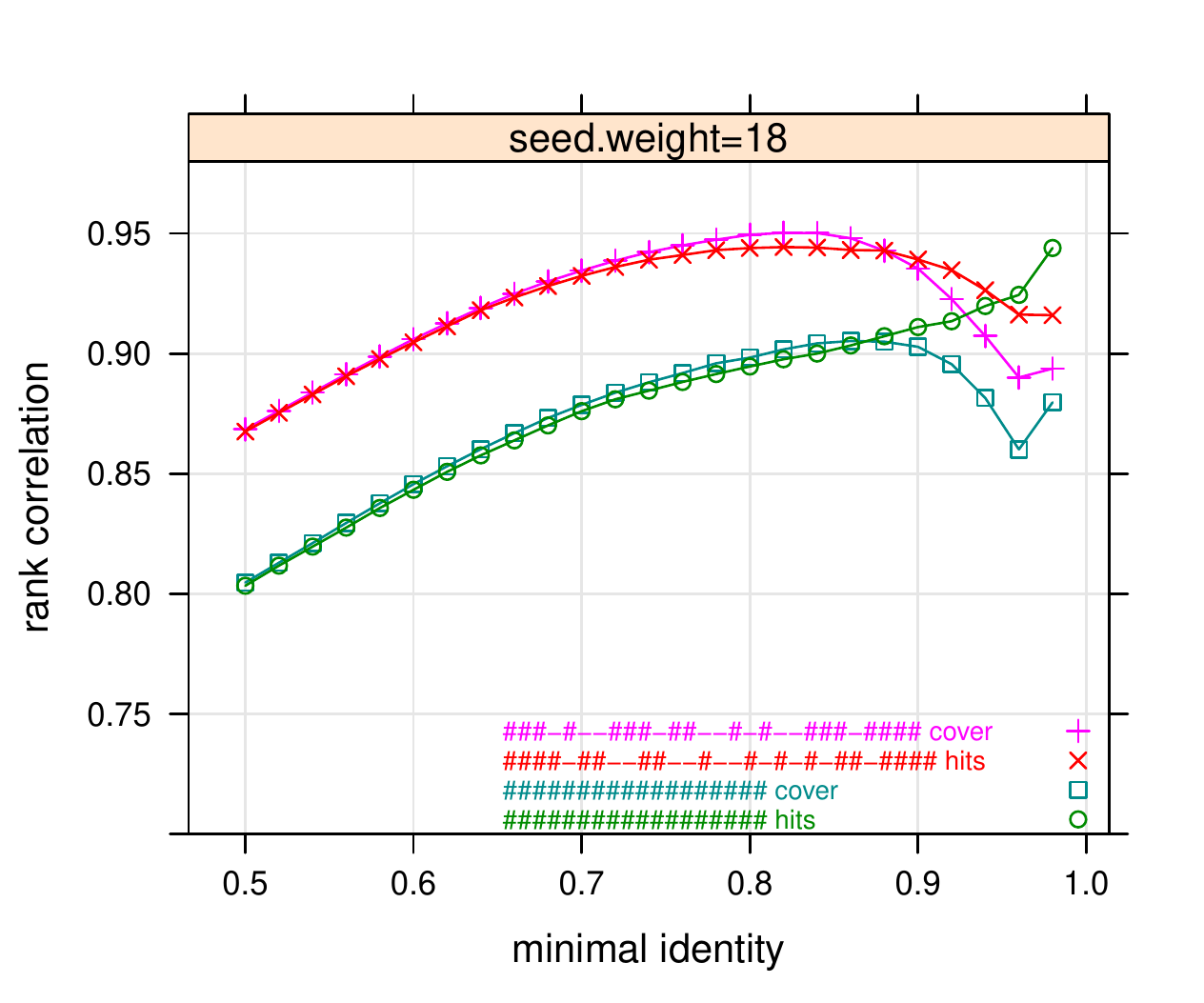}
                \label{weight18}
              \end{subfigure}\\
\begin{subfigure}[b]{0.49\textwidth}
  \includegraphics[width=\textwidth]{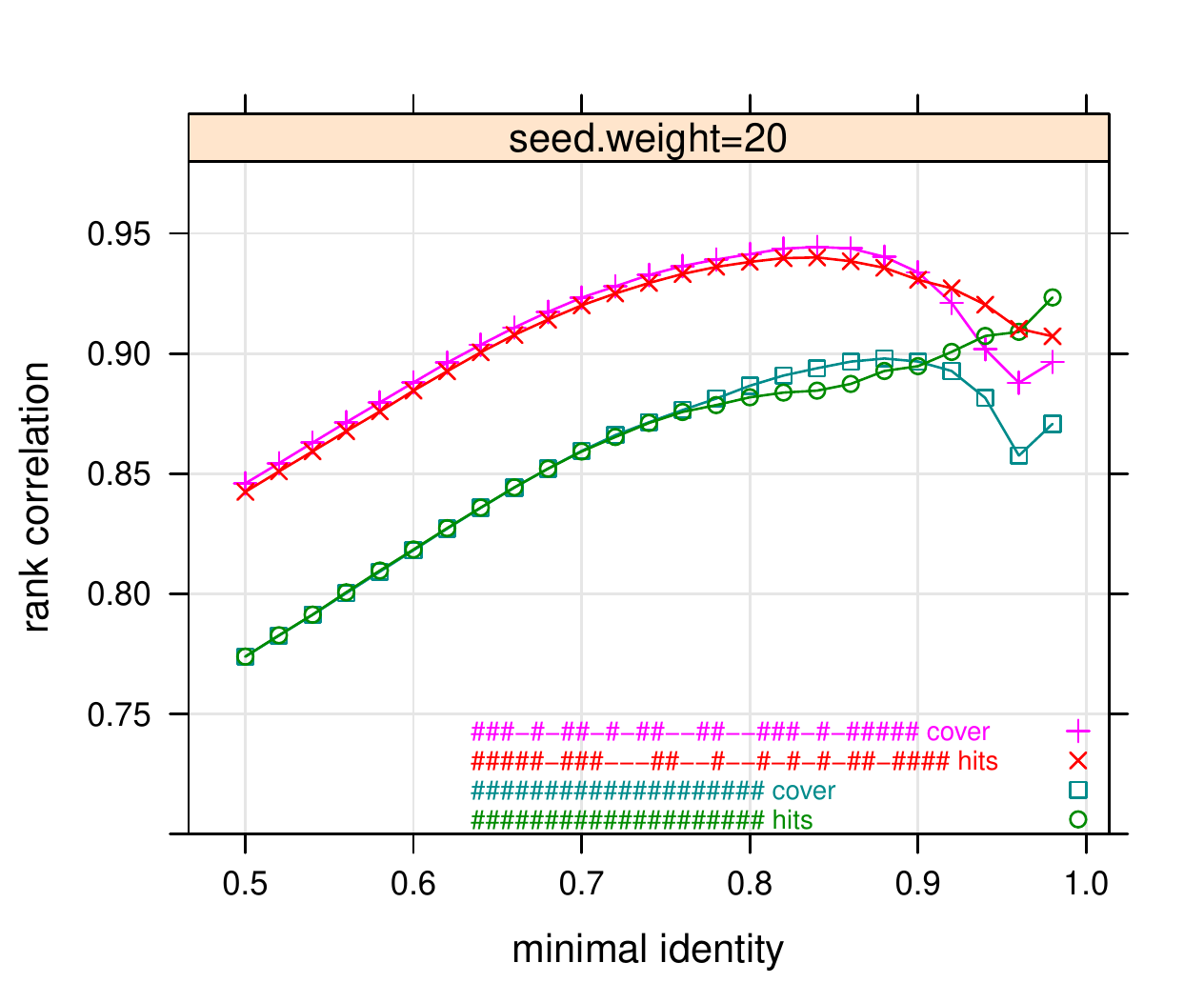}
                \label{weight20}
              \end{subfigure}
\begin{subfigure}[b]{0.49\textwidth}
  \includegraphics[width=\textwidth]{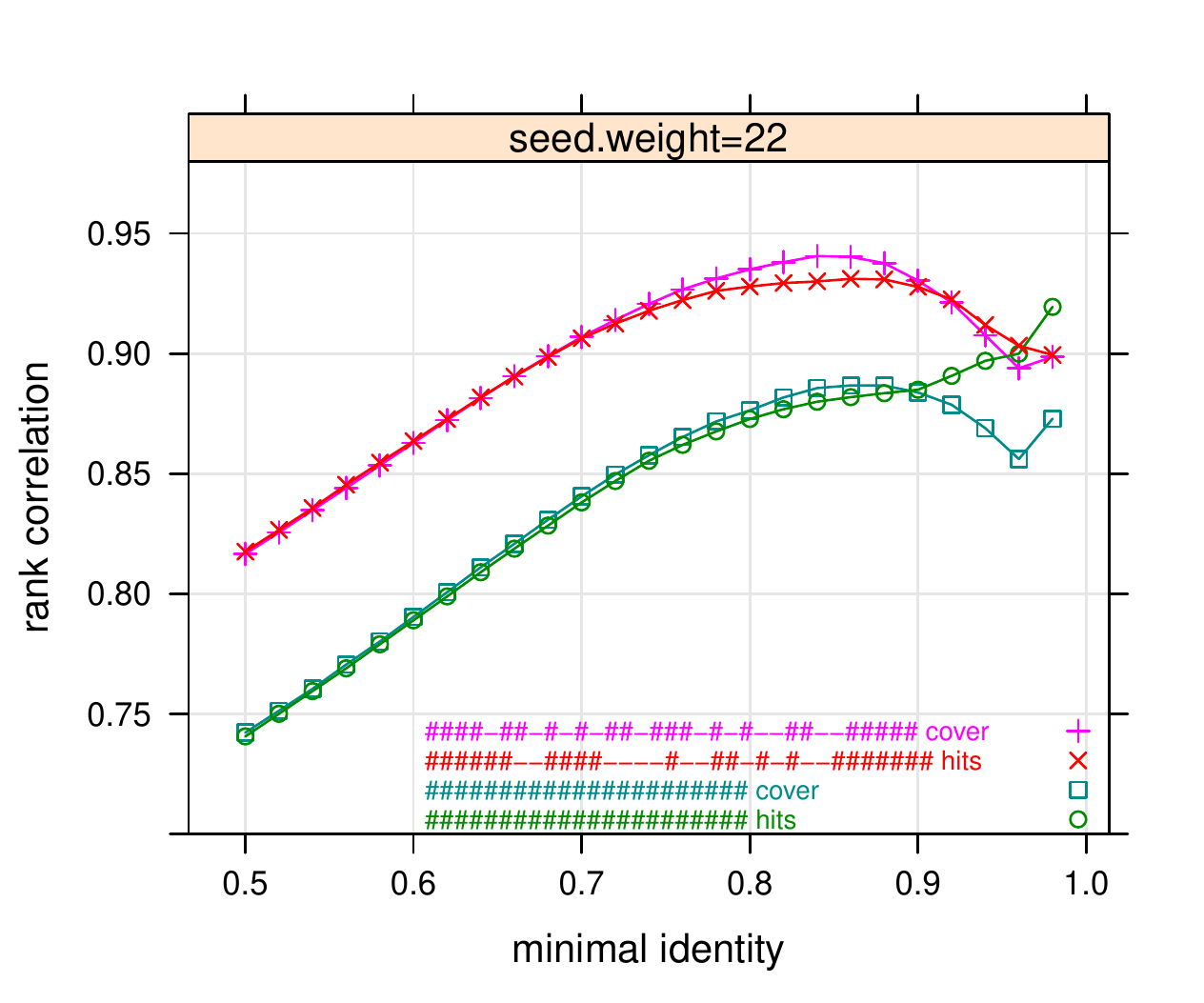}
                \label{weight22}
              \end{subfigure}\\
\begin{subfigure}[b]{0.49\textwidth}
  \includegraphics[width=\textwidth]{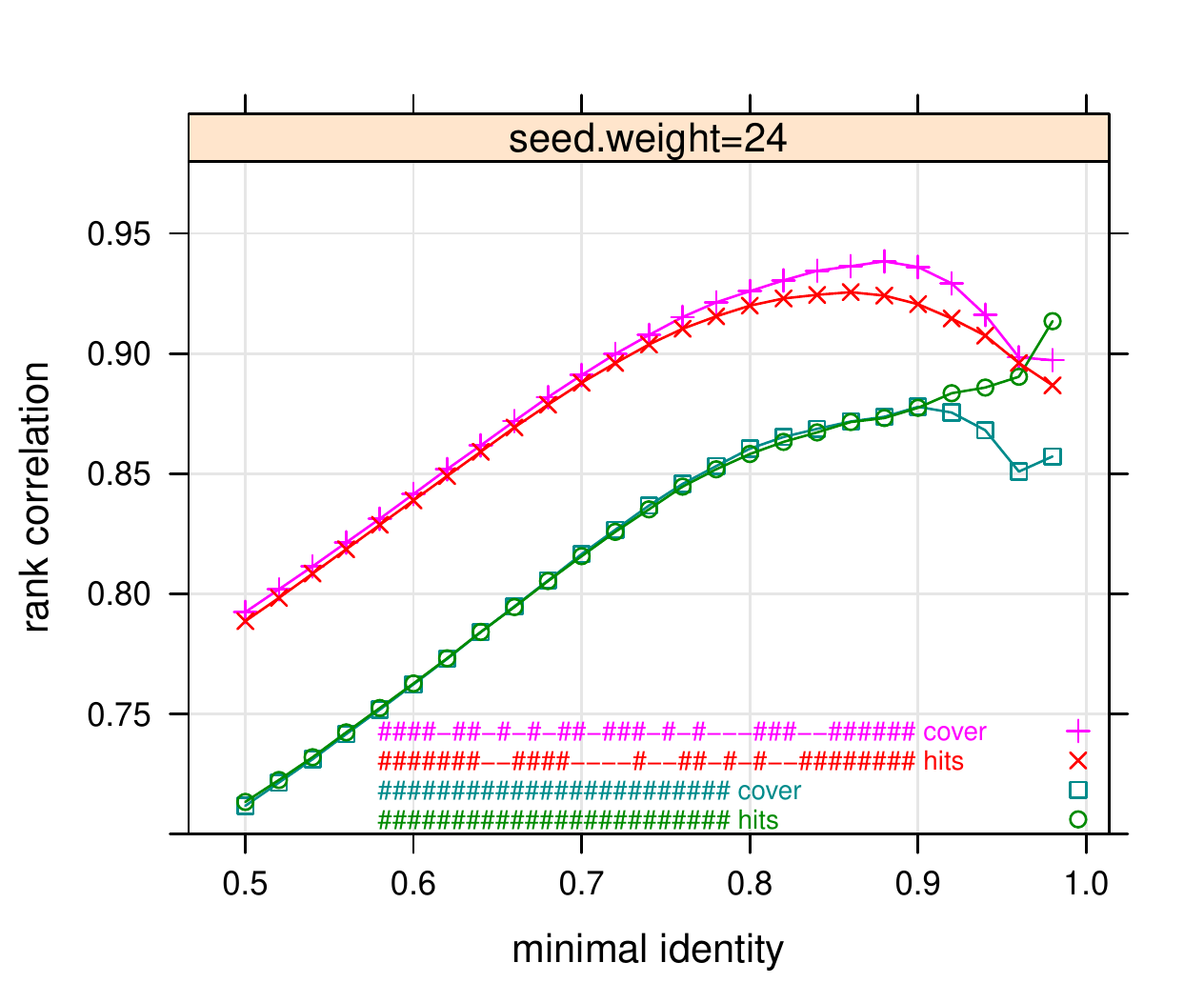}
                \label{weight24}
              \end{subfigure}
\caption{Spearman's rank correlation between score and counts, depending on
  the minimal identity rate.}\label{spearman}
\end{figure}

In metagenomic projects, reads to be classified do not necessarily
come from genomes stored in the database, but can come from genomes of
other species. These species can be genetic variants of species of the
database, such as different strains of the same bacteria, but can also
come from organisms represented in the database only at the rank of genus
or family, or may even have no representatives at all at low taxonomic
ranks. Therefore, an accurate mapping of a read to a corresponding
clade requires not just assigning it to the appropriate sampled
genome, but estimating its distances to each of the genomes in order to
locate its position within the whole taxonomic tree. 

With this motivation, we turned to the question how
well the measured counts correlate with the alignment score. For a
fixed minimal identity rate $p_{id}$, 
we randomly sampled gapless alignments of length
$100$ with identity rate from interval $[p_{id}..1]$, and
collected pairs (number of mismatches, count), where, as before,
'count' stands for either the number of hits or the coverage of a
given seed. For these data, we computed Spearman's rank correlation. 

Results are presented on plots in Figure~\ref{spearman}. 
They show that when the identity rate of alignments takes a large
range of values (minimal id rate smaller than $\approx 0.9$), 
spaced seeds yield a significantly higher correlation than
contiguous seeds, for both hit-number and coverage
counts. Furthermore, the coverage count slightly outperforms the
hit-number count, especially for spaced seeds and larger weights. 

For high-similarity alignments, however, the picture changes: the coverage
count loses its performance, with its correlation value sharply
decreasing. Furthermore, the
correlation of hit-number goes down as well for spaced seeds, while
it continues to grow for contiguous seeds ending up by reaching and
even slightly outperforming the one for spaced seed. 
This is due to a larger span of spaced seeds and to their 
combinatorial properties that cause the hit number values to be less
sharply concentrated at certain values, and therefore to be less well
correlated with the number of mismatches. 

In conclusion, while spaced seeds provide a much better
estimator for alignments whose quality ranges over a large interval, for
high-quality alignments ($>90\%$ of identity), the hit number of
contiguous seed becomes a better estimator. The superiority of
hit-number over coverage for high-quality alignments has also
been reported in \cite{pmid25393923}. Along with Spearman's
correlation, we also made an analysis of mutual information computed
on the same data (data not shown) that confirmed the above conclusions. 

\subsection{Correlation on real genomes}
\label{corr-genomes}

To validate the conclusions of the previous section in a real-life
metagenomics framework and to analyse more closely how well different
counts for a read correlate with the best alignment of the read to a
real genomic sequence, we implemented another series of experiments. 

Given a genome $G$, we generated a set of Illumina-like single-end reads 
by selecting random substrings of $G$ of length $L$ ($L=100$) and
introducing $k$ mismatch errors, with $k$ drawn randomly between $1$ and $20$ for every read.
For each read, we computed the counts -- hit number and coverage -- with respect to genome $G$ under
a given seed, similar to Section~\ref{binary-on-genomes}. Collected
data have then been analysed. 

\begin{figure}
\centering
\begin{subfigure}[b]{0.49\textwidth}
  \includegraphics[width=\textwidth]{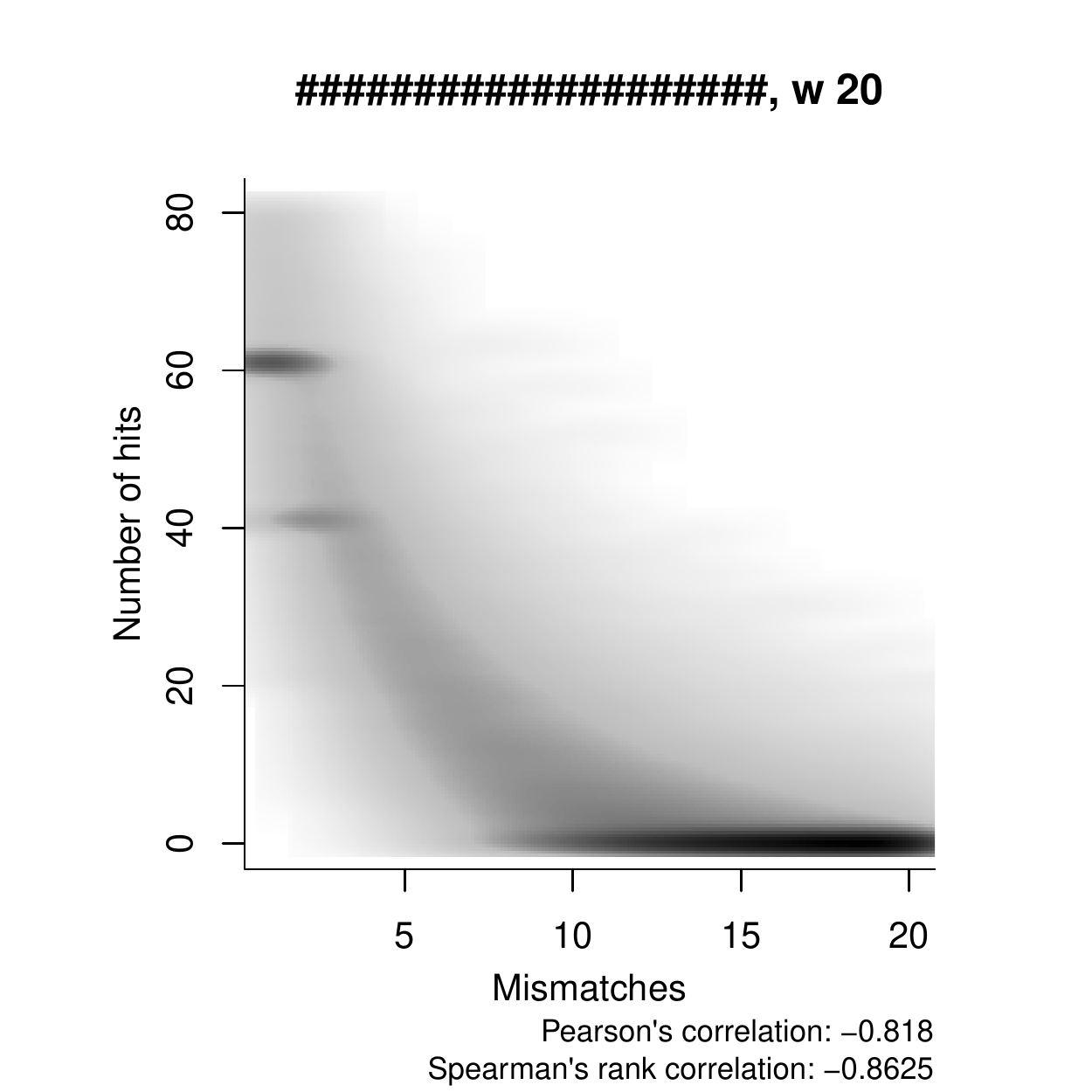}
                \label{weight-20-cont-hit-scatter}
              \end{subfigure}
\begin{subfigure}[b]{0.49\textwidth}
  \includegraphics[width=\textwidth]{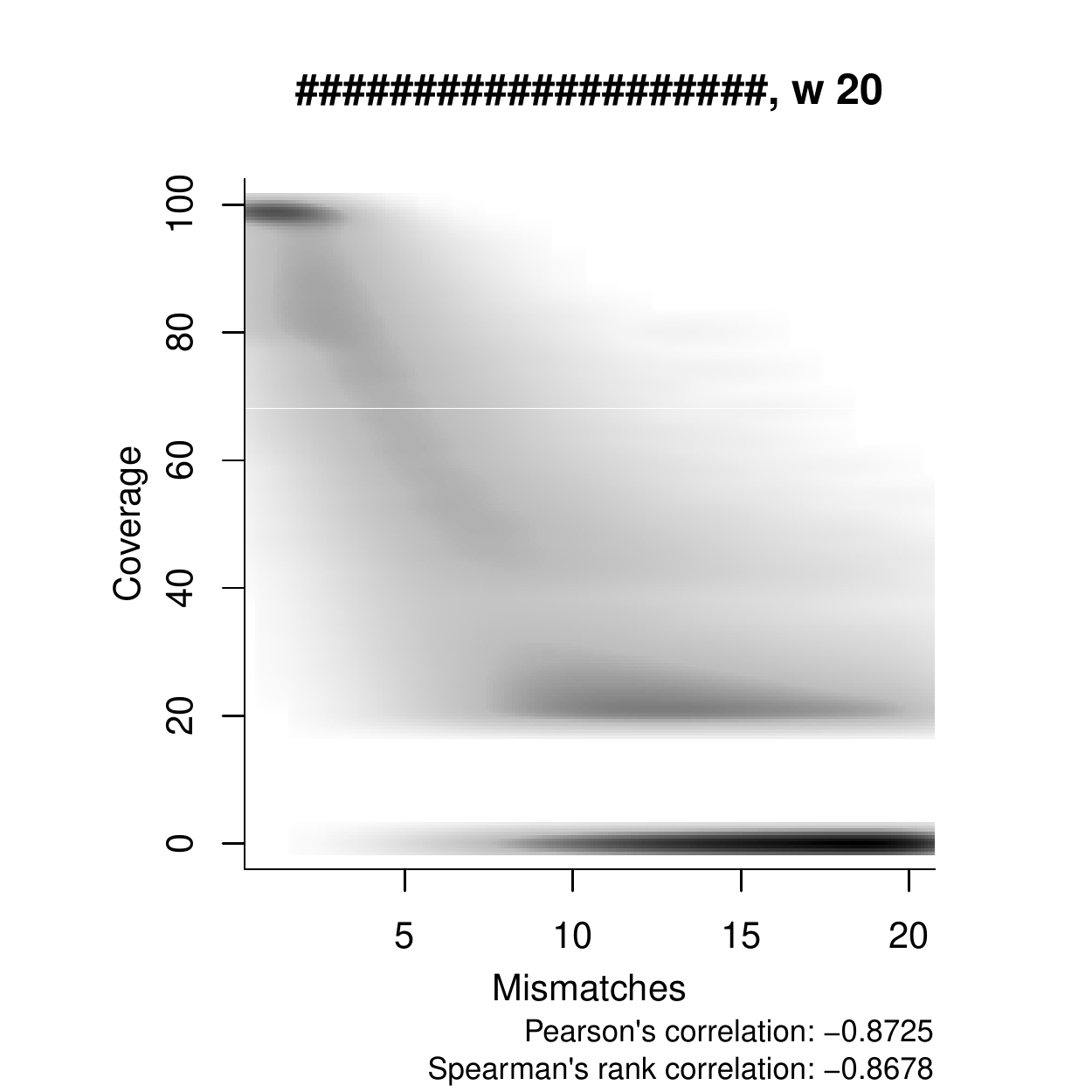}
                \label{weight-20-cont-cover-scatter}
              \end{subfigure}\\
\begin{subfigure}[b]{0.49\textwidth}
  \includegraphics[width=\textwidth]{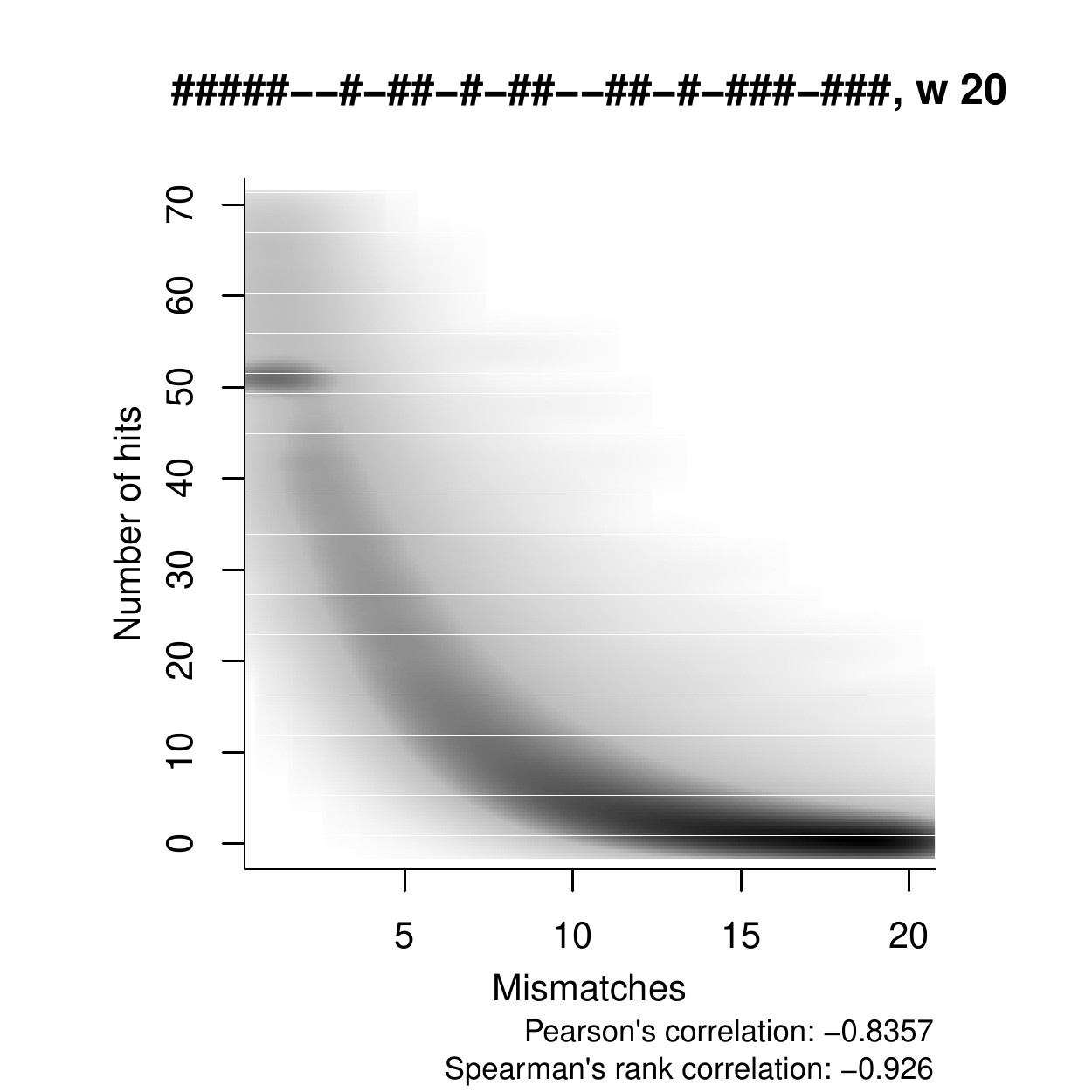}
                \label{weight-20-spaced-hit-scatter}
              \end{subfigure}
\begin{subfigure}[b]{0.49\textwidth}
  \includegraphics[width=\textwidth]{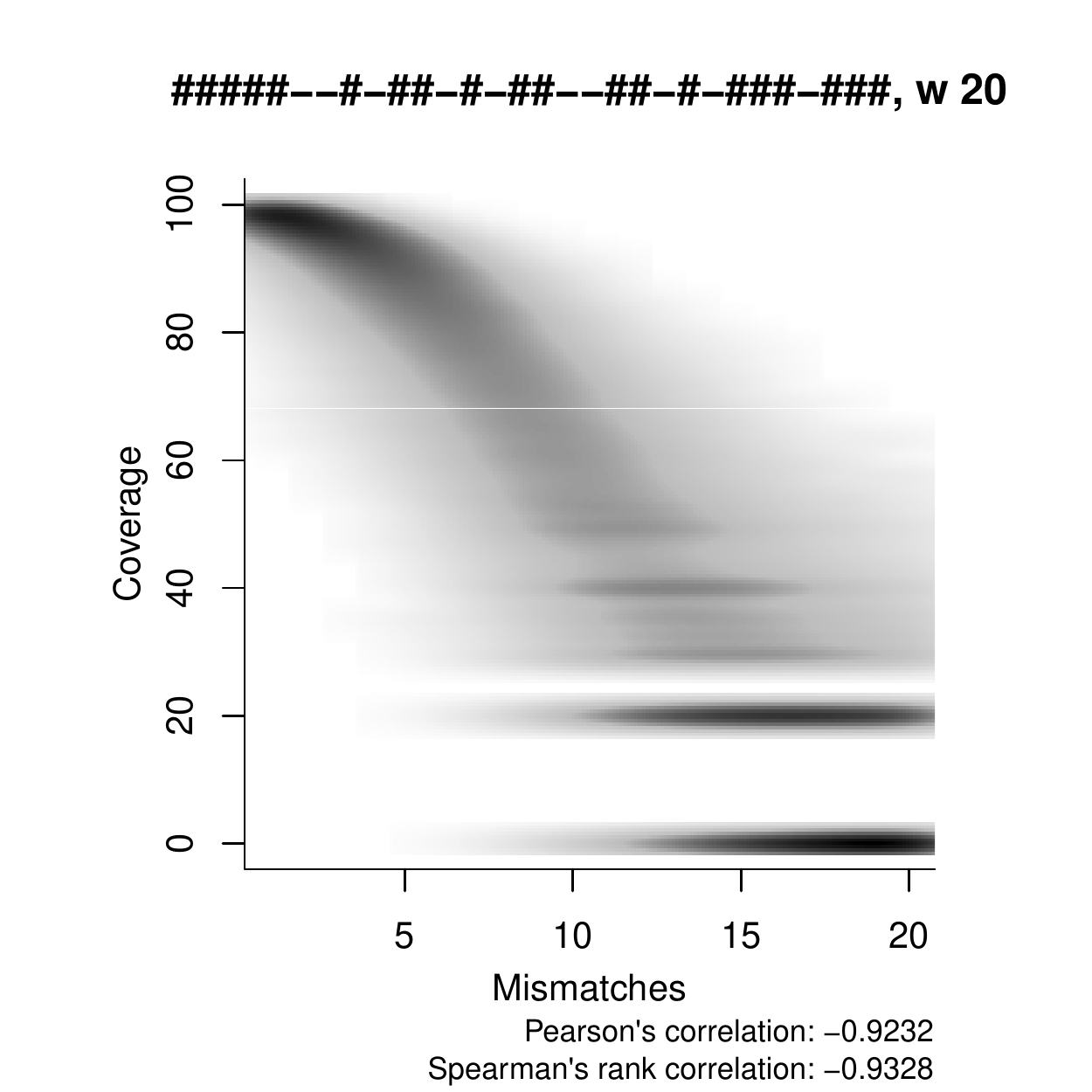}
                \label{weight-20-spaced-cover-scatter}
              \end{subfigure}
\caption{Hit number (left plots) and coverage (right plots) depending
  on the number of mismatches in randomly generated reads. Seed is
  shown above the plot, and Spearman's and Pearson's correlations are
  shown below. Grayscale shows the density of
  reads. Experiments made on {\em M.tuberculosis} 
  genome. \label{clouds}}
\end{figure}

This experimental setup has been applied to {\em Mycobacterium
  tuberculosis} genome, a typical result (seed weight 20) is shown
in Figure~\ref{clouds}. 
Each plot shows the
density of reads for each pair (number of mismatches, count),
depending on the seed (contiguous or spaced) and count (hit-number or
coverage). Spearman's and Pearson's correlation coefficients are shown
for each plotted dataset. 

\begin{figure}
\centering
\begin{subfigure}[b]{0.49\textwidth}
  \includegraphics[width=\textwidth]{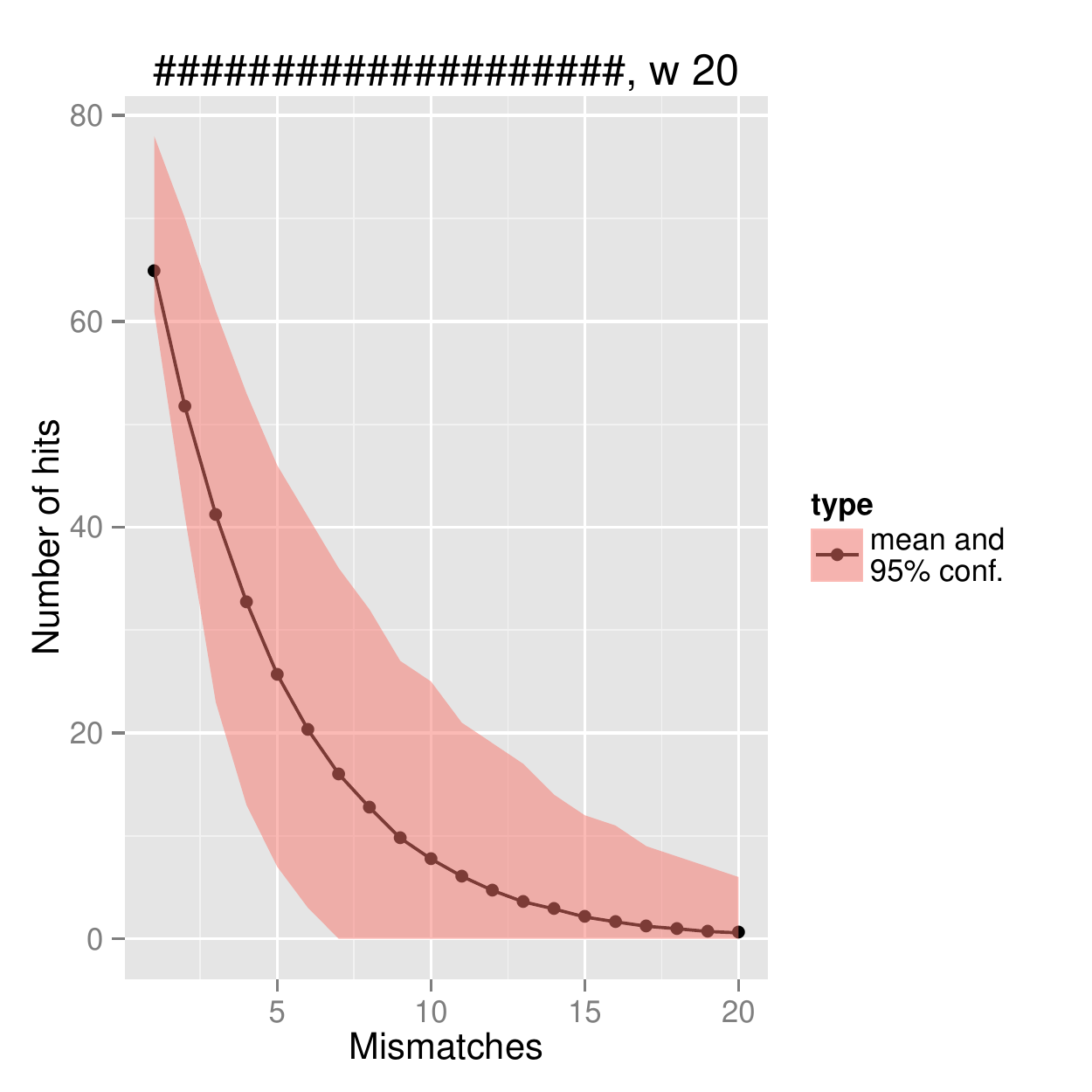}
                \label{weight-20-cont-hit-confidence}
              \end{subfigure}
\begin{subfigure}[b]{0.49\textwidth}
  \includegraphics[width=\textwidth]{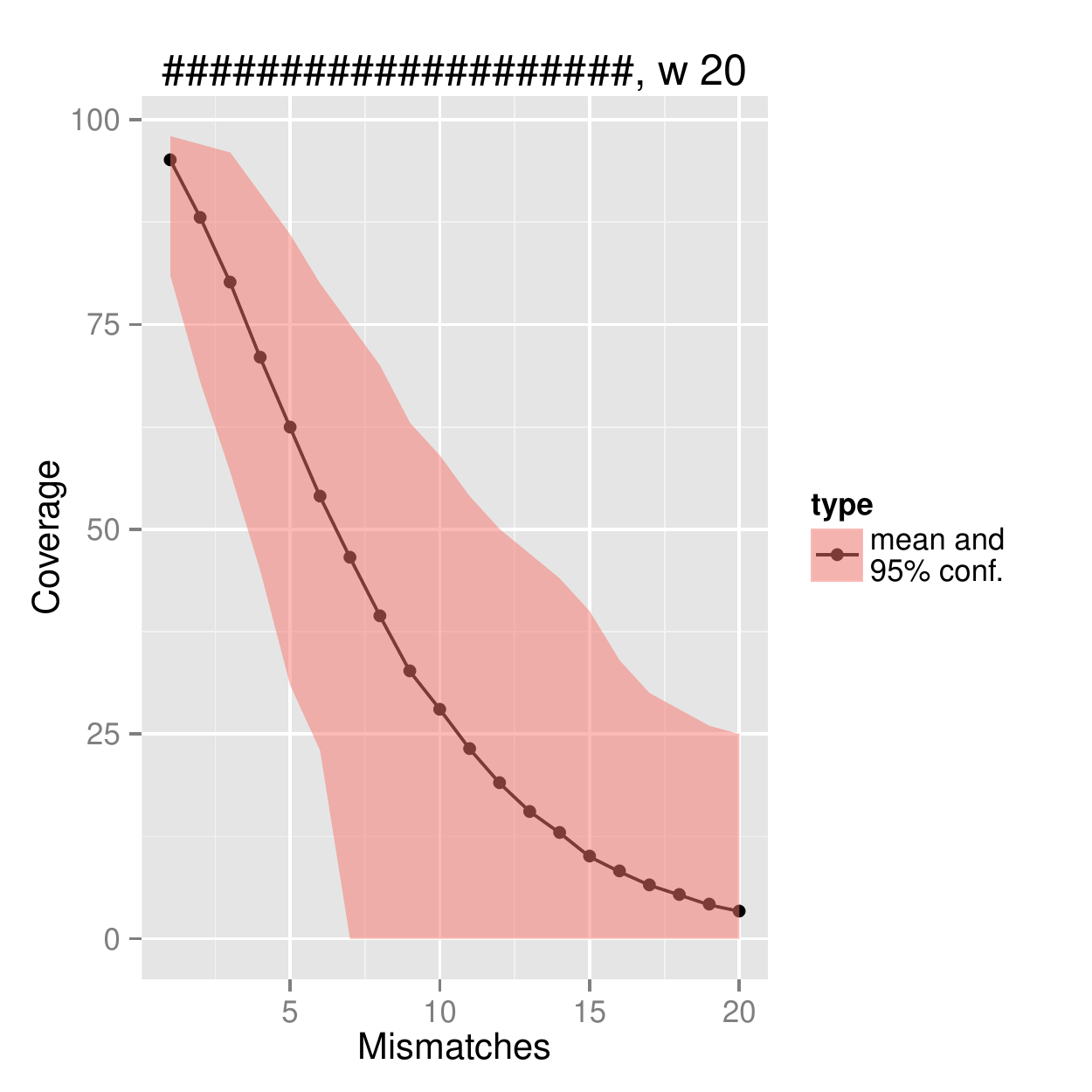}
                \label{weight-20-cont-cover-confidence}
              \end{subfigure}\\
\begin{subfigure}[b]{0.49\textwidth}
  \includegraphics[width=\textwidth]{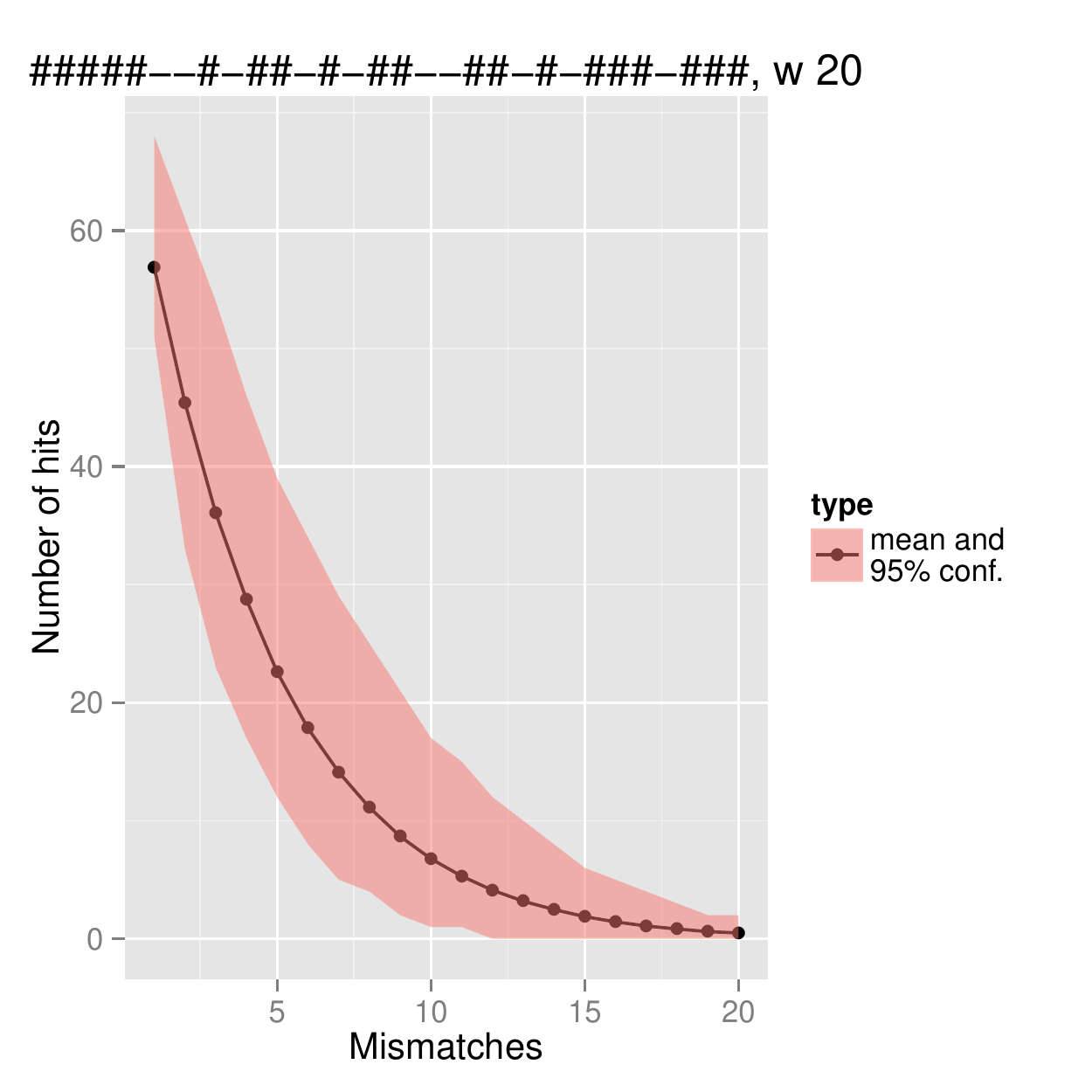}
                \label{weight-20-spaced-hit-confidence}
              \end{subfigure}
\begin{subfigure}[b]{0.49\textwidth}
  \includegraphics[width=\textwidth]{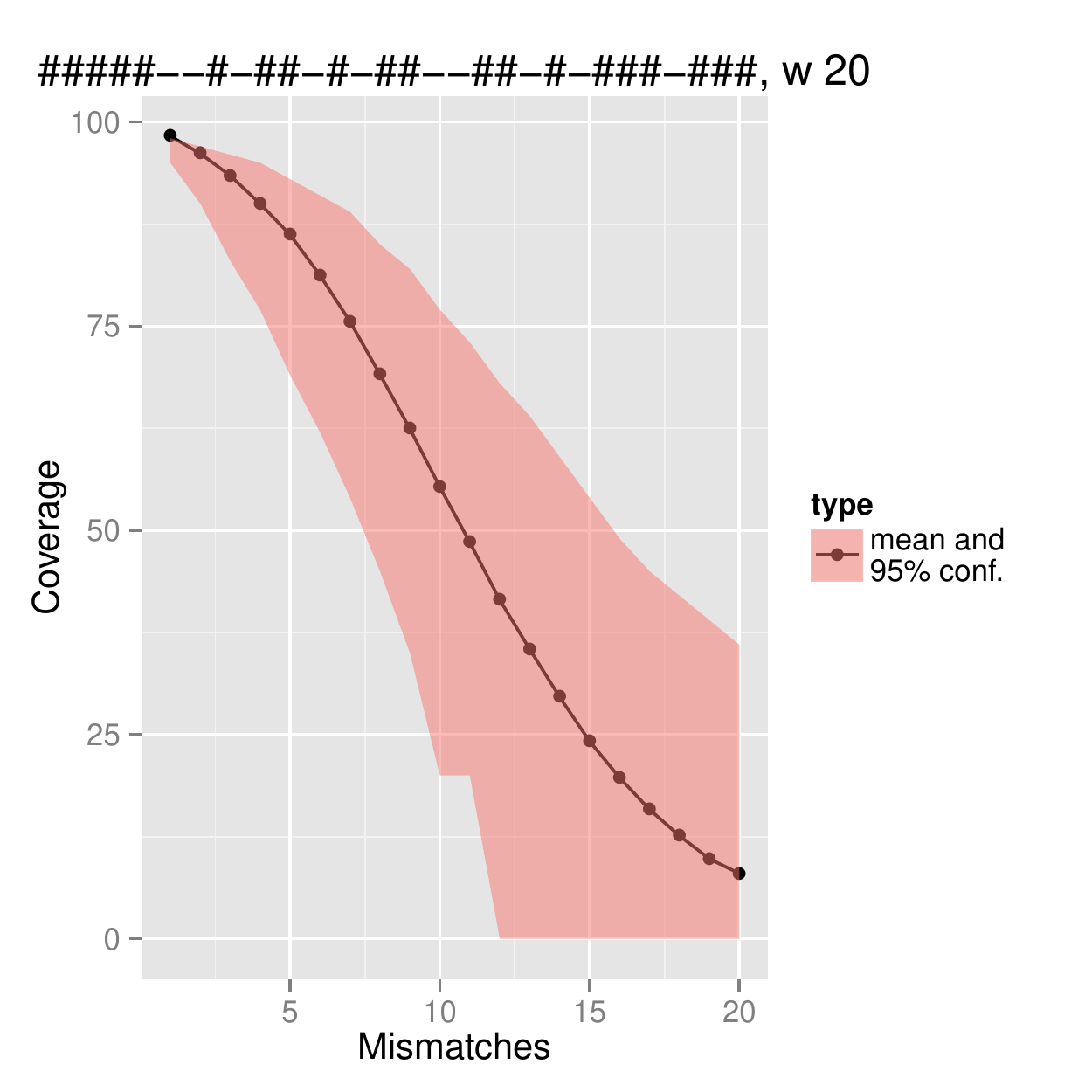}
                \label{weight-20-spaced-cover-confidence}
              \end{subfigure}
\caption{Same data as in Figure~\ref{clouds} shown with averages and
  95\% confidence intervals \label{clouds-conf}}
\end{figure}

The plots clearly illustrate the advantage of spaced seeds over
contiguous seeds for estimating the alignment quality. Plots for
contiguous seeds are more blurred whereas plots for spaced seeds
demonstrate a better correlation between the two values. This is
confirmed by the absolute values of Spearman's rank correlation coefficient
that are significantly higher for spaced seeds, indicating a better statistical
dependence. This is further illustrated in Figure~\ref{clouds-conf}
which shows the same data through the average curve and 95\%
confidence band. It confirms that spaced seeds produce a dependence
with lower deviation from the mean, compared to contiguous seeds. 

Comparing hit-number and coverage estimators, we observe that coverage
yields a slightly better Spearman correlation and a significantly
better Pearson correlation, due to a convex shape of the dependence,
compared to the more straight dependence for the coverage. 

This analysis has been done for several other bacterial genomes,
producing similar results. Plots for other genomes and other seed
weights can be found at \url{https://github.com/gregorykucherov/spaced-seeds-for-metagenomics}.

\subsection{Large-scale experiments}
\label{kraken-exp}

In order to validate the advantage of spaced seeds in large-scale
metagenomic projects, we modified  {\sc
    Kraken} software \cite{pmid24580807} to make it work with
  spaced seeds rather than with contiguous seeds only.
The limitation of this comparison is that it only allows estimating the effect of using spaced
  seeds combined with the {\sc Kraken} algorithm,  and within its
  implementation. On the other hand, 
this procedure allows us to estimate the effect of spaced seeds in an 
unbiased manner, by keeping unchanged all other factors that might
influence the results. 

Our extended implementation, that we call {\sc seed-Kraken}, allows
the user to specify a spaced seed as a parameter. 
For a set of genomes, a database of spaced $k$-mers matching the seed is constructed, 
which is later used to classify reads through the original {\sc
  Kraken} algorithm. Since {\sc Kraken} uses $k$-mer counting {\sc
  Jellyfish} program \cite{marccais2011fast} as the $k$-mer indexing
engine, we had also to modify {\sc Jellyfish} to allow it to deal with
spaced $k$-mers. 

Integration of spaced seed into {\sc Kraken} required a minor
modification of the way {\sc Kraken} deals with complementary
sequences. In {\sc Kraken}, complementary $k$-mers have a
single representative in the index, the lexicographical smallest of
the two. With spaced seeds, dealing with complementary sequencing is
more delicate, as the complement of a spaced $k$-mer does not
match the same seed but its inverse. To cope with this, we modified
{\sc Kraken} to index each distinct $k$-mer. We then processed each read
in direct and complementary directions separately and select the one
which produced more hits. Compared to original {\sc Kraken}, this
procedure takes more index space (additional $\sim 5\%$ in practice)
and doubles $k$-mer query time. 


\begin{figure}	
	\centering
        \centerline{	  
	  \includegraphics[width=1\textwidth]{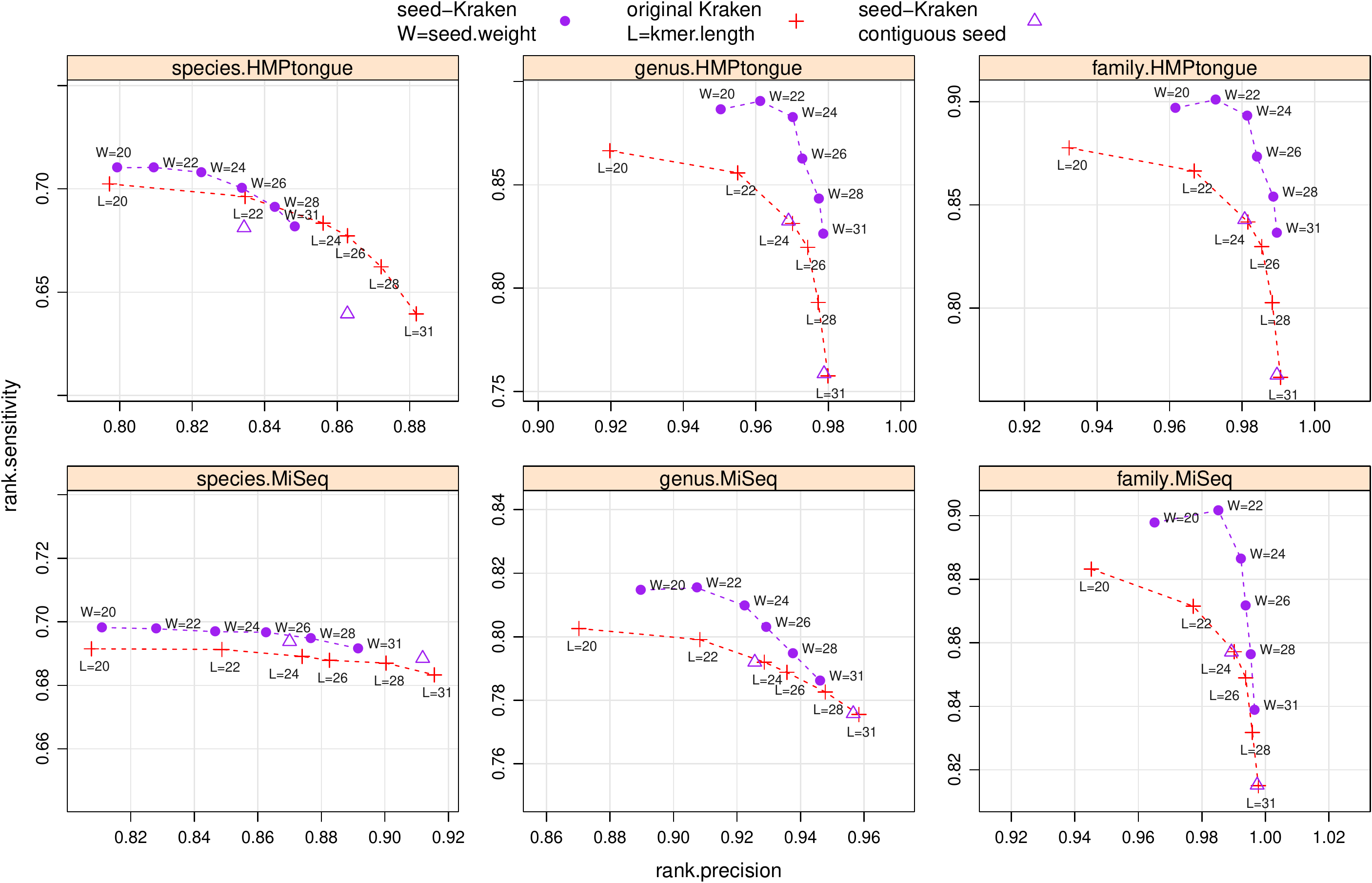}
        }                
	\caption{
Sensitivity/precision  of {\sc seed-Kraken} (circle points) and
original {\sc Kraken} (cross points) 
for HMPtongue and MiSeq datasets and three taxonomic levels: species,
genus and family.
         Triangle points correspond to {\sc seed-Kraken} 
         run on contiguous seed of weight 24 and 31, plotted to highlight the effect of the 
         change in the assignment algorithm. 
	}
	\label{fig:kraken-experiments}
\end{figure}

We compared the performance of {\sc Kraken} and {\sc seed-Kraken} on
several datasets. First, we performed experiments with three 
simulated metagenomes HiSeq, MiSeq, and simBA-5 introduced
in the original work \cite{pmid24580807}, each containing 10,000
sequences. Furthermore, we created a dataset from Human Microbiome
Project data by randomly selecting 50,000 sequences from SRS011086 Tongue
dorsum metagenomic sample \linebreak[4]
(\href{http://hmpdacc.org/HMSCP/}{http://hmpdacc.org/HMSCP/}). 
Here we only report on results for MiSeq and HMPtongue datasets and refer to
the supplementary material for a complete account including results
for HiSeq and simBA-5. MiSeq is a merge of {\sc Illumina} reads 
of 10 bacterial genomes, and HMPtongue is a random sample of real {\sc
  Illumina} whole-metagenome sequences. 

Due to resource limitations, the database we used in MiSeq experiments was half of 
the size of the {\sc Kraken}'s default database (which requires 70GB of RAM). 
Our database was obtained by choosing a
single representative strain of each bacteria species, except for the species from HiSeq and MiSeq metagenomes for which 
all strains were included. 
Overall, this represented 915 genomes of total size 3.3GB. 
For HMPtongue dataset, this database was extended with a subset of HMP
reference library, 0.8GB in total, including references for the selected
50,000 sequences. 

For each metagenomic dataset, we measured the sensitivity (percentage
of correctly classified reads out of all reads) and precision (percentage of correct
classifications out of all classifications) of {\sc Kraken} and {\sc
  seed-Kraken} at three taxonomic levels: species, genus
and family. In each case, this has been done with seeds of different
weights between 20 and 31, and for each weight, {\sc seed-Kraken} has
been run on a few different spaced seeds (see Conclusions).

Figure~\ref{fig:kraken-experiments} shows sensitivity-precision
ROC-curves (Receiver Operating Characteristic)  for {\sc seed-Kraken},
and for the unmodified {\sc Kraken}. In the case of {\sc seed-Kraken},
the ``best performing'' seed is charted for each weight. 
Furthermore, triangle points
correspond to {\sc seed-Kraken} run on contiguous seeds, plotted in order to
measure the effect of our modification in dealing with complementary
sequences. 

At the levels of genus and family, spaced seeds consistently show a
better sensitivity-precision trade-off, with the sensitivity increase of
about 2 percentage points for MiSeq and 3-5 points for HMPtongue, for
a given precision rate. The results of {\sc seed-Kraken} with contiguous
seeds (triangle points) confirm that this improvement is due to the use of spaced seeds
and not to our slight modification of the assignment algorithm due to complementary sequences. For small
weights (20-22), a spaced seed achieves simultaneously a better
sensitivity and a better precision than the contiguous seed of the
same weight. When the weight grows, the increment in precision
disappears reaching the level of the contiguous counterpart, or
sometimes coming down below it. However, this is largely compensated
by the increase in sensitivity. 

For the species level, the picture turns out to be more involved. 
Here we observe that due to the small modification of the assignment
algorithm, {\sc seed-Kraken} run with contiguous seeds (triangle
points) shows a modified behavior compared to {\sc
  Kraken}. Specifically, we observe a drop in precision and a gain in
sensitivity, and those are different for MiSeq and HMPtongue
datasets. The reason for this is that {\sc seed-Kraken} makes more
species-level classifications than {\sc Kraken} but at the same time, makes more inaccurate assignments 
to a closely related organism (typically, different strain of the same
bacteria), which eventually leads to a lower precision. This phenomenon has a
bigger impact for rich databases (HMPtongue experiment) compared to
``sparse'' databases where each species is represented by few
organisms (MiSeq). As for the contribution of spaced seeds, we observe
an improved sensitivity-precision trade-off here as well. Compared to
{\sc seed-Kraken} applied to contiguous seeds, this improvement is
small for MiSeq but significant for HMPtongue, which shows a
correction capacity of spaced seeds w.r.t. erroneous assignments to
close strains. Compared to the original {\sc Kraken}, we obtain a
sensitivity increment of about 1\% which becomes smaller (MiSeq) or
completely disappears (HMPtongue) when the seed weight grows. 

As mentioned earlier, the spaced seed corresponding to each plotted
{\sc seed-Kraken} point has been selected out of a few (usually two to
four) seeds tried. The full list of seeds applied in experiments and
corresponding results can be found in the supplementary material. Here
we just mention that for large weights (24 and more) the span of the
seed becomes an important factor, with seeds of large span showing a
drop in sensitivity and best seeds being those with relatively few jokers. 

Building a {\sc seed-Kraken} limited database takes approximately 1 hour on a server with 20 CPU cores,
and the resulting size is $26$ GB for seed of weight $24$, which compares to the $25$GB for original {\sc Kraken}.
The classification running times are longer than for original {\sc Kraken} by a factor of $3$ to $5$.

\section{Conclusions}

Through a series of computational experiments, we showed
that spaced seeds significantly improve the accuracy of metagenomic
classification of short NGS reads. The superiority of spaced seeds for
different variants of alignment-free sequence comparison has been recently demonstrated by
other authors as well
\cite{OnoderaShibuya13,pmid25033408,DBLP:journals/bioinformatics/LeimeisterBHLM14,pmid25393923}.
In this work, we specifically focused on the metagenomics setting
characterized by very large volumes of data, both in terms of the
number of reads and the size of genomic database. This quantity of
data precludes using some alignment-free comparison techniques, and
leaves room only for highly time- and space-efficient
approaches. Note also that in our setting, we have to compare short
sequences (reads) with long ones (whole genomes), which makes an
important difference with problems considered in
\cite{DBLP:journals/bioinformatics/LeimeisterBHLM14,pmid25393923,pmid25685176}. 
For example, in the framework of metagenomic classification, it is
hardly conceivable taking into account $k$-mer frequencies
\cite{DBLP:journals/bioinformatics/LeimeisterBHLM14}, as this
information would be computationally difficult to utilize. 

Another improvement considered in
\cite{DBLP:journals/bioinformatics/LeimeisterBHLM14,pmid25393923,pmid25685176}
is to use {\em multiple seeds}, i.e. several seeds simultaneously
instead of a single one. This extension is known to bring an advantage
in seed-and-extend sequence alignment
\cite{BuhlerRECOMB04,PatternHunter04}, and
\cite{DBLP:journals/bioinformatics/LeimeisterBHLM14,pmid25393923} show
that this improvement applies to alignment-free comparison as
well. However, each seed requires to build a separate index for
database genomic sequences, and therefore it appears computationally
difficult to use multiple seeds in metagenomics, unless some new
indexing techniques are designed for this purpose. 

In our work, we studied three estimators: hit number, coverage and
Jaccard index. Hit number and coverage behave similarly in
classification (Section~\ref{binary}), but Jaccard index generally improves on
them in the case of mapping to real genomes
(Section~\ref{binary-on-genomes}), due to the correction w.r.t. the
$k$-mer-richness. Considered as an estimator of alignment
quality (Section~\ref{seed-meta}, \ref{corr-genomes}), coverage
provides a certain advantage over hit number. More subtle
estimators can be considered as well, e.g. by taking into account the
position of $k$-mer in the read (reflecting the sequencing error
rate), and this provides an issue for further study.

Designing efficient seeds for metagenomic classification is another
important issue that goes beyond the present study. Note that optimal
spaced seeds for seed-and-extend alignment are generally not optimal
for alignment-free $k$-mer-based comparison \cite{pmid25393923}. In
\cite{pmid25393923}, the authors designed (sub-)optimal seeds
maximizing the Pearson correlation between hit-number/coverage count
and the alignment quality. Their solution is implemented in  {\sc
  Iedera} software\footnote{the latest version of {\sc Iedera}
  performs design  for Spearman correlation as well}
(\href{http://bioinfo.lifl.fr/yass/iedera}{http://bioinfo.lifl.fr/yass/iedera})
\cite{KucherovNoeRoytbergJBCB06}. 
On the other hand, recent work \cite{pmid25747382} introduces {\em quadratic
  residue seeds} (QR-seeds) for seed-and-extend alignment, which
present a good performance and have the advantage of easy design,
avoiding the computationally expensive enumeration of {\sc Iedera}. 
In our work, we used both {\sc Iedera} and QR-seeds adapted to our
setting. We observed that in most cases, {\sc Iedera} seeds are
superior (being designed specifically for our task) but in a few
cases, QR-seeds demonstrated equal or even better
performance\footnote{e.g. best results of Table~\ref{anth_pum_lich} for weights
  14-18 were obtained with QR-seeds}. This may be due to
their large span (cf. supplement material) for which applying {\sc
  Iedera} is computationally costly. 

\bigskip
 We now summarize the main contributions of our work. 
\begin{itemize}
\item We showed that spaced seeds can drastically improve the success
  rate of binary classification of alignments into two categories, each
  defined by a specific mismatch rate. Here the classification is done
  through ``querying'' an alignment using a seed as a mask and
  reporting whether the seed applies at a given position. For example,
  in discriminating between alignments of length 100 with mismatch
  rate $0.2$ and $0.3$, a spaced seed of weight 16 achieves 63\% of
  success while a contiguous seed of the same weight achieves only
  40\% (Table~\ref{0203}). 

Recently, spaced seeds have been shown to define more efficient
kernels for SVM classification of both protein
\cite{OnoderaShibuya13} and nucleotide (RNA) sequences
\cite{pmid25393923}.  Compared to these works, here we demonstrate the
superiority of spaced seeds in a very simple classification setting,
where sequences have to be classified according to identity rate,
without a training stage and without resorting to SVM machinery. 

\item 
We showed experimentally that spaced seeds allow for a better
  classification of NGS reads coming from a genome $G$ between two
  other genomes $G_1$ and $G_2$ of the same genus. Here reads are
  classified according to the phylogenetic distance
  between $G$ and $G_1$ and $G$ and $G_2$ respectively. We established
  that in this task, Jaccard index provides an advantage over hit-number and
  coverage which is especially important for seeds of small weight. 

\item We studied how well different estimators (coverage/hit-number
  combined with spaced/contiguous seed) correlate with the alignment
  quality, by measuring Spearman's rank correlation
  coefficient and mutual information coefficient. Here again, we
  observed a significantly better correlation produced by spaced
  seeds, but only when the alignment quality varies
  over a sufficiently large range, starting from identity rate around 0.9
  or below. On the other hand, if only high-quality alignment are
  targeted (id rate at least 0.9) then the correlation produced by
  spaced seeds becomes lower, with hit-number measure over a {\em
    contiguous} seed eventually becoming the best for alignments of id
  rate about 0.95 or more. 

\item We also measured the correlation produced on real genomes within
  the metagenomic classification approach of
  \cite{pmid23828782,pmid24580807}. For this, we assumed that the
  ``closeness'' of a read to a genome is characterized by the quality
  of the alignment (in our case, the number of mismatches), and
  computed the correlation produced by different counts on simulated
  reads. These experiments confirmed the superiority of spaced seeds
  as well. Moreover, they showed that coverage combined with spaced
  seeds provides the best option, leading to the highest Spearman's
  correlation but also to a significantly higher Pearson's
  correlation. The latter means that this estimator induces a
  dependency closer to linear, which can be a useful feature for
  classification algorithms. 

\item Finally, we compared spaced and contiguous seeds through
  large-scale metagenomics experiments. We modified {\sc Kraken} software \cite{pmid24580807} to make it work with spaced seeds,
  without modifying the core classification algorithm or any other
  parts of the software, with the only exception being the way the
  complementary sequences are dealt with. With just replacing
  contiguous seeds by spaced seeds, {\sc Kraken} showed a consistent
  improvement of specificity/sensitivity trade-off at genus and family
  levels and a dataset-dependent improvement at the species level. 
\end{itemize}

Real data experiments of
Sections~\ref{binary-on-genomes}, \ref{corr-genomes} have been done
using {\sc SnakeMake} \cite{Koster2012}. 

\bigskip
Overall, all our experiments corroborate the thesis of better performance of
spaced seeds for metagenomic classification. 
Many further questions are raised by this work. Our results remain
to be explained with more rigorous probabilistic arguments, similarly
to how it has been done for spaced seeds applied to seed-and-extend
paradigm \cite{pmid17666769}. 
While there are obvious similarities between the two applications, the
underlying ``mechanisms'' seem to be different. 
One sign of this difference is that optimal spaced seeds for the two
problems are not the same, as mentioned earlier. 
%

Experiments with {\sc Kraken} (Section~\ref{kraken-exp}) give a strong
evidence that spaced seeds can improve the classification accuracy
in real-life large-scale metagenomic projects. One further improvement
would be to implement coverage and Jaccard measures that showed, 
in general,
a better performance compared to the hit number. Introducing spaced
seeds rises new issues, such as the construction of an efficient index
of the database, or adapting the algorithm of computing the most
likely node of the taxonomic tree from counts produced by individual
genomes, i.e. leaves of the tree. These questions are a subject for
future work. 

\paragraph{Acknowledgements.} All authors have been supported by the
ABS4NGS grant and by Labex B\'ezout of the French government (program \emph{Investissement
  d'Avenir}). Many thanks to Michal Ziv-Ukelson for discussions that
sparked our interest to this research and to Laurent No\'e for help
with seed design. 

\bibliography{biblio}

\begin{thebibliography}{10}

\bibitem{pmid20504335}
L.~D. Alcaraz, G.~Moreno-Hagelsieb, et~al.
\newblock {{U}nderstanding the evolutionary relationships and major traits of
  {B}acillus through comparative genomics}.
\newblock {\em BMC Genomics}, 11:332, 2010.

\bibitem{GBLAST97}
S.~Altschul, T.~Madden, et~al.
\newblock Gapped {BLAST }and {PSI-BLAST}: a new generation of protein database
  search programs.
\newblock {\em Nucleic Acids Research}, 25(17):3389--3402, 1997.

\bibitem{pmid23828782}
S.~K. Ames, D.~A. Hysom, et~al.
\newblock {{S}calable metagenomic taxonomy classification using a reference
  genome database}.
\newblock {\em Bioinformatics}, 29(18):2253--2260, Sep 2013.

\bibitem{pmid21304684}
A.~F. Auch, M.~von Jan, et~al.
\newblock {{D}igital {D}{N}{A}-{D}{N}{A} hybridization for microbial species
  delineation by means of genome-to-genome sequence comparison}.
\newblock {\em Stand Genomic Sci}, 2(1):117--134, 2010.

\bibitem{pmid21810899}
E.~Bao, T.~Jiang, I.~Kaloshian, and T.~Girke.
\newblock {{S}{E}{E}{D}: efficient clustering of next-generation sequences}.
\newblock {\em Bioinformatics}, 27(18):2502--2509, Sep 2011.

\bibitem{pmid18974822}
A.~Ben-Hur, C.~S. Ong, et~al.
\newblock {{S}upport vector machines and kernels for computational biology}.
\newblock {\em PLoS Comput. Biol.}, 4(10):e1000173, Oct 2008.

\bibitem{DBLP:conf/spire/BensonM08}
G.~Benson and D.~Mak.
\newblock Exact distribution of a spaced seed statistic for {DNA} homology
  detection.
\newblock In {\em String Processing and Information Retrieval, 15th
  International Symposium, {SPIRE} 2008, Melbourne, Australia, November 10-12,
  2008. Proceedings}, volume 5280 of {\em Lecture Notes in Computer Science},
  pages 282--293. Springer, 2008.

\bibitem{pmid21527926}
A.~Brady and S.~Salzberg.
\newblock {{P}hymm{B}{L} expanded: confidence scores, custom databases,
  parallelization and more}.
\newblock {\em Nat. Methods}, 8(5):367, May 2011.

\bibitem{BrownBA08}
D.~G. Brown.
\newblock {\em Bioinformatics Algorithms: Techniques and Applications}, chapter
  A survey of seeding for sequence alignment, pages 126--152.
\newblock Wiley-Interscience (I. M{\v{a}}ndoiu, A. Zelikovsky), Fev. 2008.

\bibitem{BurkhardtKarkkainen03}
S.~Burkhardt and J.~K{\"a}rkk{\"a}inen.
\newblock Better filtering with gapped $q$-grams.
\newblock {\em Fundamenta Informaticae}, 56(1-2):51--70, 2003.
\newblock Preliminary version in Combinatorial Pattern Matching 2001.

\bibitem{pmid25747382}
L.~Egidi and G.~Manzini.
\newblock {{M}ultiple seeds sensitivity using a single seed with threshold}.
\newblock {\em J Bioinform Comput Biol}, page 1550011, Feb 2015.

\bibitem{pmid25033408}
M.~Ghandi, D.~Lee, et~al.
\newblock {{E}nhanced regulatory sequence prediction using gapped k-mer
  features}.
\newblock {\em PLoS Comput. Biol.}, 10(7):e1003711, Jul 2014.

\bibitem{pmid10831447}
E.~Helgason, O.~A. Okstad, et~al.
\newblock {{B}acillus anthracis, {B}acillus cereus, and {B}acillus
  thuringiensis--one species on the basis of genetic evidence}.
\newblock {\em Appl. Environ. Microbiol.}, 66(6):2627--2630, Jun 2000.

\bibitem{pmid21690186}
D.~H. Huson, S.~Mitra, et~al.
\newblock {{I}ntegrative analysis of environmental sequences using
  {M}{E}{G}{A}{N}4}.
\newblock {\em Genome Res.}, 21(9):1552--1560, Sep 2011.

\bibitem{pmid22028628}
E.~Karsenti, S.~G. Acinas, P.~Bork, C.~Bowler, C.~De~Vargas, J.~Raes,
  M.~Sullivan, D.~Arendt, F.~Benzoni, J.~M. Claverie, M.~Follows, G.~Gorsky,
  P.~Hingamp, D.~Iudicone, O.~Jaillon, S.~Kandels-Lewis, U.~Krzic, F.~Not,
  H.~Ogata, S.~Pesant, E.~G. Reynaud, C.~Sardet, M.~E. Sieracki, S.~Speich,
  D.~Velayoudon, J.~Weissenbach, and P.~Wincker.
\newblock {{A} holistic approach to marine eco-systems biology}.
\newblock {\em PLoS Biol.}, 9(10):e1001177, Oct 2011.

\bibitem{pmid25884504}
J.~Kawulok and S.~Deorowicz.
\newblock {{C}o{M}eta: {C}lassification of {M}etagenomes {U}sing k-mers}.
\newblock {\em PLoS ONE}, 10(4):e0121453, 2015.

\bibitem{BLAT02}
W.~J. Kent.
\newblock {BLAT}--the {BLAST}-like alignment tool.
\newblock {\em Genome Research}, 12:656--664, 2002.

\bibitem{Koster2012}
J.~K\"{o}ster and S.~Rahmann.
\newblock Snakemake - a scalable bioinformatics workflow engine.
\newblock {\em Bioinformatics}, 28(19):2520--2522, 2012.

\bibitem{KucherovNoeRoytbergJBCB06}
G.~Kucherov, L.~No{\'e}, et~al.
\newblock A unifying framework for seed sensitivity and its application to
  subset seeds.
\newblock {\em Journal of Bioinformatics and Computational Biology},
  4(2):553--569, November 2006.

\bibitem{pmid19261174}
B.~Langmead, C.~Trapnell, et~al.
\newblock {{U}ltrafast and memory-efficient alignment of short {D}{N}{A}
  sequences to the human genome}.
\newblock {\em Genome Biol.}, 10(3):R25, 2009.

\bibitem{DBLP:journals/bioinformatics/LeimeisterBHLM14}
C.-A. Leimeister, M.~Boden, et~al.
\newblock Fast alignment-free sequence comparison using spaced-word
  frequencies.
\newblock {\em Bioinformatics}, 30(14):1991--1999, 2014.

\bibitem{pmid19451168}
H.~Li and R.~Durbin.
\newblock {{F}ast and accurate short read alignment with {B}urrows-{W}heeler
  transform}.
\newblock {\em Bioinformatics}, 25(14):1754--1760, Jul 2009.

\bibitem{PatternHunter04}
M.~Li, B.~Ma, et~al.
\newblock {P}attern{H}unter {II}: Highly sensitive and fast homology search.
\newblock {\em Journal of Bioinformatics and Computational Biology},
  2(3):417--439, 2004.
\newblock Earlier version in GIW 2003.

\bibitem{LindgreenEtAl15}
S.~Lindgreen, K.~L. Adair, and P.~Gardner.
\newblock An evaluation of the accuracy and speed of metagenome analysis tools.
\newblock {\em bioRxiv}, 2015.

\bibitem{PatternHunter02}
B.~Ma, J.~Tromp, et~al.
\newblock {P}attern{H}unter: Faster and more sensitive homology search.
\newblock {\em Bioinformatics}, 18(3):440--445, 2002.

\bibitem{pmid22962338}
S.~S. Mande, M.~H. Mohammed, et~al.
\newblock {{C}lassification of metagenomic sequences: methods and challenges}.
\newblock {\em Brief. Bioinformatics}, 13(6):669--681, Nov 2012.

\bibitem{marccais2011fast}
G.~Mar{\c{c}}ais and C.~Kingsford.
\newblock A fast, lock-free approach for efficient parallel counting of
  occurrences of k-mers.
\newblock {\em Bioinformatics}, 27(6):764--770, 2011.

\bibitem{pmid23103880}
S.~Marco-Sola, M.~Sammeth, et~al.
\newblock {{T}he {G}{E}{M} mapper: fast, accurate and versatile alignment by
  filtration}.
\newblock {\em Nat. Methods}, 9(12):1185--1188, Dec 2012.

\bibitem{pmid25685176}
B.~Morgenstern, B.~Zhu, et~al.
\newblock {{E}stimating evolutionary distances between genomic sequences from
  spaced-word matches}.
\newblock {\em Algorithms Mol Biol}, 10:5, 2015.

\bibitem{NatMeth09}
{Nath Meth Editorial}.
\newblock Metagenomics versus {M}oore's law.
\newblock {\em Nat Meth}, 6(9):623--623, 09 2009.

\bibitem{NoeKucherovBMC04}
L.~No{\'e} and G.~Kucherov.
\newblock Improved hit criteria for {DNA} local alignment.
\newblock {\em BMC Bioinformatics}, 5(149), 14 October 2004.

\bibitem{pmid25393923}
L.~No{\'e} and D.~E. Martin.
\newblock {{A} {C}overage {C}riterion for {S}paced {S}eeds and {I}ts
  {A}pplications to {S}upport {V}ector {M}achine {S}tring {K}ernels and k-{M}er
  {D}istances}.
\newblock {\em J. Comput. Biol.}, 21(12):947--963, Dec 2014.

\bibitem{OnoderaShibuya13}
T.~Onodera and T.~Shibuya.
\newblock The gapped spectrum kernel for support vector machines.
\newblock In P.~Perner, editor, {\em Machine Learning and Data Mining in
  Pattern Recognition}, volume 7988 of {\em Lecture Notes in Computer Science},
  pages 1--15. Springer Berlin Heidelberg, 2013.

\bibitem{pmid25879410}
R.~Ounit, S.~Wanamaker, T.~J. Close, and S.~Lonardi.
\newblock {{C}{L}{A}{R}{K}: fast and accurate classification of metagenomic and
  genomic sequences using discriminative k-mers}.
\newblock {\em BMC Genomics}, 16:236, 2015.

\bibitem{pmid19819907}
J.~Peterson, S.~Garges, et~al.
\newblock {{T}he {N}{I}{H} {H}uman {M}icrobiome {P}roject}.
\newblock {\em Genome Res.}, 19(12):2317--2323, Dec 2009.

\bibitem{QinEtAlNature10}
J.~Qin, R.~Li, J.~Raes, M.~Arumugam, K.~S. Burgdorf, C.~Manichanh, T.~Nielsen,
  N.~Pons, F.~Levenez, T.~Yamada, D.~R. Mende, J.~Li, J.~Xu, S.~Li, D.~Li,
  J.~Cao, B.~Wang, H.~Liang, H.~Zheng, Y.~Xie, J.~Tap, P.~Lepage, M.~Bertalan,
  J.-M. Batto, T.~Hansen, D.~Le~Paslier, A.~Linneberg, H.~B. Nielsen,
  E.~Pelletier, P.~Renault, T.~Sicheritz-Ponten, K.~Turner, H.~Zhu, C.~Yu,
  S.~Li, M.~Jian, Y.~Zhou, Y.~Li, X.~Zhang, S.~Li, N.~Qin, H.~Yang, J.~Wang,
  S.~Brunak, J.~Dore, F.~Guarner, K.~Kristiansen, O.~Pedersen, J.~Parkhill,
  J.~Weissenbach, P.~Bork, S.~D. Ehrlich, and J.~Wang.
\newblock A human gut microbial gene catalogue established by metagenomic
  sequencing.
\newblock {\em Nature}, 464(7285):59--65, 03 2010.

\bibitem{BLASTZ03}
S.~Schwartz, J.~Kent, A.~Smit, Z.~Zhang, R.~Baertsch, R.~Hardison, D.~Haussler,
  and W.~Miller.
\newblock Human--mouse alignments with {BLASTZ}.
\newblock {\em Genome Research}, 13:103--107, 2003.

\bibitem{BuhlerRECOMB04}
Y.~Sun and J.~Buhler.
\newblock Designing multiple simultaneous seeds for {DNA} similarity search.
\newblock In {\em Proceedings of the 8th Annual International Conference on
  Computational Molecular Biology (RECOMB04), San Diego (California)}. ACM
  Press, March 2004.

\bibitem{pmid22966151}
H.~Teeling and F.~O. Glockner.
\newblock {{C}urrent opportunities and challenges in microbial metagenome
  analysis--a bioinformatic perspective}.
\newblock {\em Brief. Bioinformatics}, 13(6):728--742, Nov 2012.

\bibitem{pmid21684354}
E.~Tortoli.
\newblock {{P}hylogeny of the genus {M}ycobacterium: many doubts, few
  certainties}.
\newblock {\em Infect. Genet. Evol.}, 12(4):827--831, Jun 2012.

\bibitem{pmid15001713}
J.~C. Venter, K.~Remington, J.~F. Heidelberg, A.~L. Halpern, D.~Rusch, J.~A.
  Eisen, D.~Wu, I.~Paulsen, K.~E. Nelson, W.~Nelson, D.~E. Fouts, S.~Levy,
  A.~H. Knap, M.~W. Lomas, K.~Nealson, O.~White, J.~Peterson, J.~Hoffman,
  R.~Parsons, H.~Baden-Tillson, C.~Pfannkoch, Y.~H. Rogers, and H.~O. Smith.
\newblock {{E}nvironmental genome shotgun sequencing of the {S}argasso {S}ea}.
\newblock {\em Science}, 304(5667):66--74, Apr 2004.

\bibitem{pmid24819825}
S.~Vinga.
\newblock {{E}ditorial: {A}lignment-free methods in computational biology}.
\newblock {\em Brief. Bioinformatics}, 15(3):341--342, May 2014.

\bibitem{VogelEtAlNatRevMicro09}
T.~M. Vogel, P.~Simonet, J.~K. Jansson, P.~R. Hirsch, J.~M. Tiedje, J.~D. van
  Elsas, M.~J. Bailey, R.~Nalin, and L.~Philippot.
\newblock Terragenome: a consortium for the sequencing of a soil metagenome.
\newblock {\em Nat Rev Micro}, 7(4):252--252, 04 2009.

\bibitem{pmid24580807}
D.~E. Wood and S.~L. Salzberg.
\newblock {{K}raken: ultrafast metagenomic sequence classification using exact
  alignments}.
\newblock {\em Genome Biol.}, 15(3):R46, 2014.

\bibitem{pmid17666769}
L.~Zhang.
\newblock {{S}uperiority of spaced seeds for homology search}.
\newblock {\em IEEE/ACM Trans Comput Biol Bioinform}, 4(3):496--505, 2007.

\end{thebibliography}

\end{document}